\documentclass[useAMS,usenatbib]{mn2e}
\usepackage[dvips]{epsfig}
\title[Shear-driven instabilities in the solar interior]{Interactions between
  magnetohydrodynamic shear instabilities and convective flows in
  the solar interior}
\author[L.\ J.\ Silvers, P.\ J.\ Bushby  $\&$  M.\ R.\ E.\ Proctor ]{L.\ J.\ Silvers $^{1}$\thanks{E-mail:
ljs53@damtp.cam.ac.uk (LJS)}, P.\ J.\ Bushby$^{2}$ $\&$ M.\ R.\ E.\ Proctor$^{1}$\\
$^{1}$Department of Applied Mathematics and Theoretical
  Physics, University of Cambridge, Cambridge, CB3 OWA, United Kingdom\\
  $^{2}$School of Mathematics and Statistics, Newcastle University,
  Newcastle Upon Tyne, NE1 7RU, United Kingdom\\}
\begin{document}

\date{Submitted ????}

\pagerange{\pageref{firstpage}--\pageref{lastpage}}
\pubyear{2009}

\maketitle

\label{firstpage}

\begin{abstract}
\noindent Motivated by the interface model for the solar dynamo, this
paper explores the complex magnetohydrodynamic interactions between
convective flows and shear-driven instabilities. Initially, we
consider the dynamics of a forced shear flow across a
convectively-stable polytropic layer, in the presence of a vertical
magnetic field. When the imposed magnetic field is weak, the dynamics
are dominated by a shear flow (Kelvin-Helmholtz type) instability. For stronger fields,
a magnetic buoyancy instability is preferred. If this stably
stratified shear layer lies below a convectively unstable region,
these two regions can interact. Once again, when the imposed field is
very weak, the dynamical effects of the magnetic field are negligible and the
interactions between the shear layer and the convective layer are
relatively minor. However, if the magnetic field is strong enough to
favour magnetic buoyancy instabilities in the shear layer, extended
magnetic flux concentrations form and rise into the convective
layer. These magnetic structures have a highly disruptive effect upon
the convective motions in the upper layer.
\end{abstract}


\section{Introduction}

The 11 year solar magnetic cycle is driven by a hydromagnetic
dynamo. However, the exact nature of this dynamo mechanism is still
not fully understood, and there are several scenarios that seek to
explain the observed behaviour. The well-known ``interface'' dynamo
model \citep{Parker}, is based on the idea that the dynamo operates in
a region that straddles the base of the solar convection zone and the
stably stratified region that lies beneath \citep[for some recent
  reviews see][]{Ossendrijver, mp06, Dormy, silvers08}. Although this
  is a conceptually appealing model for the solar dynamo, the only
  numerical investigations of the interface dynamo have been based
  upon mean-field dynamo theory \citep[see, for
  example,][]{Char,Chan,Zhang,Bushby}. In mean-field theory, several
  aspects of the dynamo model (particularly the effects of turbulent
  convection) are parametrised. However, the resulting coefficients are poorly determined by both theory and observations. Due to the
computational costs involved, it has not yet been possible to
demonstrate the operation of the interface dynamo by carrying out
three-dimensional simulations of compressible
magnetohydrodynamics. Given these computational constraints, it makes
sense to investigate different components of the interface dynamo in
isolation. 

\par An important feature of the region below the solar convection
zone is the solar tachocline \citep[see][and references
therein]{SZ,CT}, which takes the form of an intense radial gradient of
the solar differential rotation. At the heart of the interface dynamo scenario is the idea that weak poloidal
magnetic fields can be amplified by the intense shears in the
tachocline, leading to the production of strong toroidal (azmimuthal)
magnetic fields. In the standard interface dynamo model, these
poloidal magnetic fields are produced in the convection zone, and are
pumped down into the tachocline by the fluid motions
\citep{TBCT1,TBC2}. In flux transport dynamo models, these poloidal
fields are transported from the surface (to the tachocline region) by
a meridional circulation \citep[see, e.g.,][]{DG09}. Wherever the
poloidal field is generated, a mechanism is needed to produce flux
structures that rise through the convection zone to the surface, where
they emerge to form active regions. The most natural mechanism for
inducing this vertical transport is magnetic buoyancy
\citep{Parker_buoy,Newcomb1961}. Strong coherent fields exert a
magnetic pressure that leads to these magnetic regions becoming less
dense than their surroundings. Provided that the ambient medium is not
too stably stratified, instabilities can occur that would appear to
allow strong fields to rise into the convection zone above. A full
discussion of magnetic buoyancy and its importance in relation to
tachocline dynamics can be found in \cite{Hughes2007}, while the role
of the tachocline in the solar cycle is described by \cite{TW07}.   

\par Until recently, most studies have addressed the
evolution of magnetic buoyancy instabilities in a prescribed layer of
magnetised fluid \citep[see, for example,][]{Catthughes,MHP,
  Wissink,fan, Evy}. However, it is not immediately obvious that a
realistic velocity shear can produce strong enough magnetic fields to
become unstable to buoyancy modes, particularly in the very
stably-stratified tachocline. To become buoyant the fields must exert
strong Lorentz forces, which will also retard the flow and resist the
field amplification.  The linear evolution of magnetic buoyancy
instabilities in a compressible magnetic layer, with an
aligned velocity shear, was considered by \citet{HT2}. They found that
magnetic buoyancy instabilities tended to be stabilised by a strong
velocity shear. Recent numerical calculations have started to address
the more complex problem of the nonlinear evolution of shear-driven
magnetic buoyancy instabilities \citep[see, for
example,][]{BCC,CBC,Vasila,Vasilb}.        

\par Using a combination of high resolution numerical simulations and
analytical calculations, \citet{Vasila,Vasilb} investigated the
stability of a magnetic layer that is generated by the action of a
strong vertical velocity shear upon an imposed uniform
magnetic field. They argued that no magnetic buoyancy instability would be
possible, in the stably-stratified tachocline, unless the magnitude of
the velocity shear were many orders of magnitude larger than the
inferred radial shear. This would have profound consequences for
the hydrodynamic stability of the shear. Defining the Richardson
number, $Ri$, to be the square of the Brunt-V\"ais\"al\"a frequency
divided by the square of the velocity gradient, a necessary
condition for hydrodynamic stability is that $Ri > 1/4$
\citep[see, for example,][]{Chand}. In the tachocline, the Richardson
number is estimated to be many orders of magnitude larger than that
given by this stability bound, which implies that the shear is
stable. However, if the velocity shear is strong enough that the
stability condition is not satisfied, as in the calculations of
\citet{Vasila,Vasilb}, then this system will be subject to shear
instabilities (of ``Kelvin-Helmholtz'' type). Clearly the situation
becomes more complicated in the  presence of an imposed magnetic
field, and the subsequent evolution depends crucially (and highly
non-trivially) upon the strength of this magnetic field. This is an
interesting problem in its own right. \citet{HT1} considered the
linear evolution of magnetised shear instabilities, whilst the
nonlinear problem has also been studied in unstratified compressible fluids
\citep{frank,ryu,Palotti} as well as in isothermal stratified layers
\citep{Bruggen}. The recent review article by \citet{GC07} describes
global magnetohydrodynamic shear instabilities in the
tachocline.

\par Although the hydrodynamic stability of the velocity shear was discussed by
\citet{Vasila,Vasilb}, the most important idea in their work was the
suggestion that shear-driven magnetic buoyancy instabilities can only
occur at very small values of the Richardson number. Since this would
appear to be incompatible with the tachocline, this would have dire
consequences for solar dynamo models. However, these results
are not conclusive. Firstly, their calculations were all performed
using a fixed value for the imposed magnetic field strength. This is
clearly an important parameter, since the Lorentz force plays a
crucial dynamical role. More importantly, recent calculations by
\citet{siletal09} have confirmed, as already known for the onset of magnetic buoyancy without shear \citep{Hughes2007}, that the onset of magnetic buoyancy
instabilities depends upon the ratio of the magnetic to thermal
diffusivities. At high Reynolds numbers, \citet{siletal09} have shown
that magnetic buoyancy instabilities can be excited with a weaker
(hydrodynamically-stable) shear if the thermal diffusivity is much
greater than the magnetic diffusivity (something that was not the case
in the original calculations of \citet{Vasila}). This is a more
encouraging result from the point of view of the solar dynamo,
although more work remains to be done.    

\par Clearly, the parametric dependence of the instabilities of a
forced shear flow, in the presence of a magnetic field, is still not
fully understood. One of the aims of this paper is to enhance our
understanding of these instabilities via a partial exploration of
parameter space. In particular, we focus attention upon the effects of
varying the strength of the imposed magnetic field (although
some variations in other parameters are also considered). Once the evolution of this system has been studied in a single convectively-stable layer
of compressible fluid, we move on to consider a more complicated
``composite'' model, which combines the stably-stratified shear layer
with an overlying convectively-unstable region. This composite model
enables us to address the interesting question, so far largely
unexplored, of how buoyant magnetic flux might interact with the fluid
in the lower convection zone. In order to limit the
computational expense of this parametric survey, we choose a stronger
velocity shear than that considered by \citet{siletal09}. This enables
us to drive buoyancy instabilities at lower Reynolds numbers, which
means that fully resolved numerical simulations can be carried out
with a coarser numerical grid. Although our chosen flow is hydrodynamically 
 unstable, it is still much weaker than the
target shear that was considered by \citet{Vasila,Vasilb}, being
mildly subsonic as opposed to highly supersonic (though still
much stronger than that found within the tachocline). Although we are
not exploring the rather extreme parameter regime that is directly
relevant for the tachocline, our choice of parameters allows us to
enhance our basic understanding of the interactions between magnetic
buoyancy and convective instabilities. Future research (which will
rely upon this work) will focus upon these phenomena at higher
Richardson numbers.  

\par The plan of the paper is as follows: In the next section we describe the set up of the model problem. In section 3, we present numerical
results from this model, describing the interactions between magnetic and hydrodynamic instabilities in a single stably-stratified polytropic
layer. In the following section, we describe the (more complicated) problem of shear-driven instabilities in the composite model. Finally,
in section 5, we conclude with a discussion of the astrophysical significance of these results.

\section[]{The model}

The  model we use is similar to that of several previous
studies of magnetoconvection \citep[see, for
example,][]{MPW,BH,Lin}. We consider the evolution of a plane
layer of electrically-conducting fluid, which is heated from below, in
the presence of a magnetic field that is initially uniform and vertical. Accordingly, we adopt a Cartesian frame of reference such that
the $z$-axis points vertically downwards, parallel to the constant
gravitational acceleration, $g$. Defining $d$ to be some
characteristic lengthscale (e.g. the depth of the convection zone in
the composite model), this fluid occupies the region $0 \le x, y
\le \lambda d$ and $0\le z \le dz_0$. We set $z_0=1$ in the single
layer calculations that are described in the next section (in which
case $d$ corresponds to the layer depth), whilst $z_0=3$ in the
composite model. Varying the parameter $\lambda$ enables us to change
the width of the domain without altering the value of $d$. In contrast
to most
previous studies of magnetoconvection \citep[although see the recent
paper by][]{Kap}, we investigate the evolution of this system under
the influence of a forced horizontal shear flow in the $x$-direction. 

\par Throughout this paper, we assume that the shear viscosity, $\mu$,
the magnetic diffusivity, $\eta$, the permeability of free space,
$\mu_0$, and the specific heat capacities at constant pressure and
density ($c_P$ and $c_V$ respectively) are all constant properties of the fluid. The thermal conductivity, $K(z)$, is assumed to be
a function of $z$. Defining $\rho$ to be the fluid density, $T$ to be the
temperature, $\mathbf{u}$ to be the fluid velocity and $\mathbf{B}$ to
be the magnetic field, the governing equations for the evolution of
this compressible fluid are given by:
\begin{eqnarray}
\label{equation1}
\frac{\partial \rho}{\partial t} + \nabla \cdot \left(\rho
  \mathbf{u}\right) &=& 0\\
\label{equation2}
\rho \left[\frac{\partial \mathbf{u}}{\partial t} + \left(\mathbf{u}\cdot\nabla\right)\mathbf{u}\right] &=& -\nabla P +
\rho g\mathbf{\hat{z}} - \mu \nabla^2 \left[U_0(z)\mathbf{\hat{x}}\right]\\ \nonumber &+& \frac{1}{\mu_0}\left(\nabla \times
  \mathbf{B}\right)\times\mathbf{B} +\nabla \cdot \left(\mu
  \mbox{\boldmath $\tau$}\right)\\
\label{equation3}
\rho c_V\left[\frac{\partial T}{\partial t} +
  \left(\mathbf{u}\cdot\nabla\right)T\right] &=&  -P\nabla \cdot
\mathbf{u} + \nabla \cdot\left[K(z)\nabla T\right] \\ \nonumber &+& \frac{\eta|\nabla\times \mathbf{B}|^2}{\mu_0} +
\frac{\mu \tau^2}{2}\\
\label{equation4}
\frac{\partial \mathbf{B}}{\partial t} &=& \nabla \times
\left[\mathbf{u}\times\mathbf{B} - \eta \nabla \times
  \mathbf{B}\right] \\
\label{equation5}
\nabla \cdot \mathbf{B}&=&0,
\end{eqnarray}
\noindent where the pressure $P$ satisfies the perfect gas law
\begin{equation}
P=R_*\rho T,\label{equation6}
\end{equation}

\noindent (defining $R_*$ to be the gas constant) and the components
of the viscous stress tensor {\boldmath $\tau$} satisfy
\begin{equation}
\tau_{ij} = \frac{\partial u_i}{\partial x_j} +\frac{\partial
  u_j}{\partial x_i} - \frac{2}{3}\frac{\partial u_k}{\partial
  x_k}\delta_{ij}.\label{equation7}
\end{equation}
\noindent Finally the scalar quantity, $U_0(z)$, represents the
horizontal shear flow. The corresponding forcing term in
Equation~\ref{equation2} ensures that any imposed shear of this form
is a solution of the horizontal component of the momentum equation (in the
absence of any other motions).
\par The boundary conditions for these
variables are consistent with those of an idealised model. All
variables are assumed to satisfy periodic boundary conditions in the $x$ and $y$ directions. The upper and lower bounding surfaces (at $z=0$
and $z=dz_0$ respectively), are assumed to be impermeable and
stress-free, and it also assumed that the magnetic field is vertical
at these boundaries. The upper boundary is held at fixed temperature,
whilst the heat flux passing through the lower surface is assumed to be
constant. This implies that:
\begin{eqnarray}
\label{equation8}
u_z= \frac{\partial u_x}{\partial z}= \frac{\partial u_y}{\partial z}=
B_x=B_y=0, \hspace{0.05in} T=T_0 \hspace{0.05in}\mbox{at $z=0$},\hspace{0.15in}\\
\label{equation9}
u_z= \frac{\partial u_x}{\partial z}= \frac{\partial u_y}{\partial z}=
B_x=B_y=0, \hspace{0.05in} \frac{\partial T}{\partial z}=C \hspace{0.05in}\mbox{at $z=dz_0$},
\end{eqnarray}
\noindent where $C$ is a constant that will depend upon the initial
conditions of the model. Note that the choice of a stress-free
boundary condition for $u_x$ implies that the imposed shear,
$\partial{U_0(z)}/\partial z$, should also be zero at these surfaces. 

These equations can be expressed in non-dimensional form, using
the scalings described by \citet{MPW}. Lengths are scaled by $d$, whilst the density and temperature are scaled by
their initial values at the top of the layer ($\rho_0$ and $T_0$
respectively). Velocities are scaled in terms of the isothermal sound
speed at the upper surface, $\sqrt{R_*T_0}$, which suggests a natural
scaling for time of $d/\sqrt{R_*T_0}$. Magnetic fields are scaled in
terms of the strength of the initial vertical magnetic field
$B_0$. Finally, we define $K_0$ to be the value of $K(z)$ at the upper
surface. When these scalings are substituted into the governing
equations, we obtain several non-dimensional parameters
that are essentially identical to those described by \citet{MPW}. These
include the dimensionless thermal diffusivity, $\kappa=K_0/d \rho_{0} c_{P}
\sqrt{(R_{*} T_{0})}$, the ratio of specific heats, $\gamma=c_P/c_V$,
the Prandtl number, $\sigma=\mu c_P/K_0$, and the ratio of the magnetic to
the thermal diffusivity at the top of the layer, $\zeta_0=\eta c_P
\rho_0/K_0$. Finally, $F=B_0^2/R_*T_0\rho_0\mu_0$ is
the ratio of the squared Alfv\'en speed to the square of the
isothermal sound speed at the top of the layer. This parameter
determines the dynamical influence of any imposed magnetic field.

With appropriate choices for these non-dimensional parameters, we
solve the equations numerically using a parallel hybrid
finite-difference/pseudo-spectral code. In this code, time-stepping is
carried out with an explicit third-order Adams-Bashforth
scheme. Horizontal derivatives are evaluated in Fourier space, whilst
vertical derivatives are calculated using fourth-order finite
differences (upwinded derivatives being used for the advective
terms). In order to carry out these simulations, grid resolutions of
$128\times128\times200$ mesh points were used for the single layer
cases, whilst $256\times256\times300$ mesh points were used for the
composite model. Some calculations were also carried out at lower
spatial resolution and comparisons of the different resolutions show
that the instabilities and structures that emerge are physical and not
artefacts of the discretization.

\section{The single layer}

In this section, we consider the evolution of this system with the
simplest possible initial configuration. An understanding of this
system will help us to interpret the results from the next
section, which deals with a much more complicated model
problem. Throughout this section, we define the computational domain
by setting $\lambda=4$ and $z_0=1$. Recalling that all lengths are
scaled in terms of $d$, this implies that $0\le x,y\le 4$ and $0\le
z\le 1$.

\subsection{Parameters}

The behaviour of this model depends crucially upon the initial
conditions that are imposed. The simplest non-trivial case to choose is that of a polytropic layer with
a constant thermal conductivity, $K(z)=K_0$. In this case, the initial
conditions are completely determined by two non-dimensional
parameters, namely the (dimensionless) temperature difference between
the upper and lower boundaries, $\theta$, and the polytropic index,
$m=gd/R_*T_0\theta$. Neglecting the effects of viscous heating, it is
straightforward to show that the governing equations have the
following (dimensionless) equilibrium solution: $T=(1+\theta z)$,
$\rho=(1+\theta z)^m $, $B_z=1$, $u_x = U_0(z)$,
$u_y=u_z=B_x=B_y=0$. Of course the effects of viscous heating will lead to a departure from this equilibrium, but we have verified (by
direct calculation) that the departure is
 negligible over the time-scales
that are considered in this paper. Therefore the above  ``equilibrium''
solution (together with a small, random, thermal perturbation) is used
as an initial condition for all the simulations that are described in
this section. Note that these initial conditions imply that the lower
boundary condition for temperature (see Equation~\ref{equation9})
becomes $\partial T/\partial z = \theta$ at $z=1$.

\begin{table}
\begin{center}
\begin{tabular}{|c|c|c|}
\hline
Param.   &   Description &  Value  \\
\hline
$\sigma$  &  Prandtl  Number  & $0.05$ \\
$m$   &  Polytropic  Index &  $1.6$  \\
$\theta$  &  Thermal  Stratification  & $0.5$ \\
$\gamma$  &  Ratio  of  Specific  Heats & $5/3$ \\
$\kappa$  & Thermal Diffusivity & $0.01$   \\
$F$ & Magnetic Field Strength & Variable \\
$\zeta_0$ & Magnetic Diffusivity & $0.2$\\
\hline
\end{tabular}
\caption{Parameter values for the single layer calculations.} \label{table1}
\end{center}
\end{table}

\begin{figure}
\begin{center}
\epsfig{file=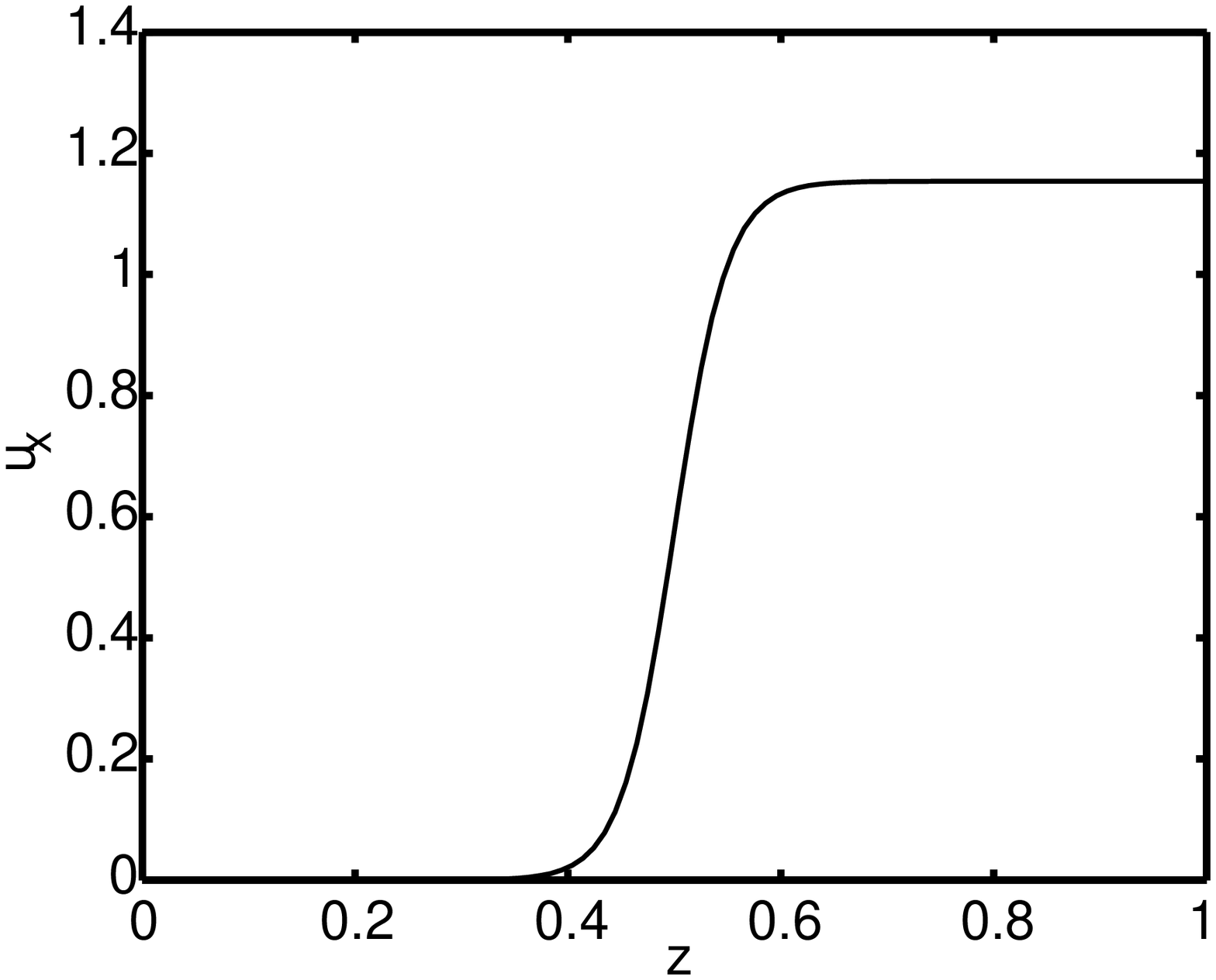,width=8cm, height=5cm}
\epsfig{file=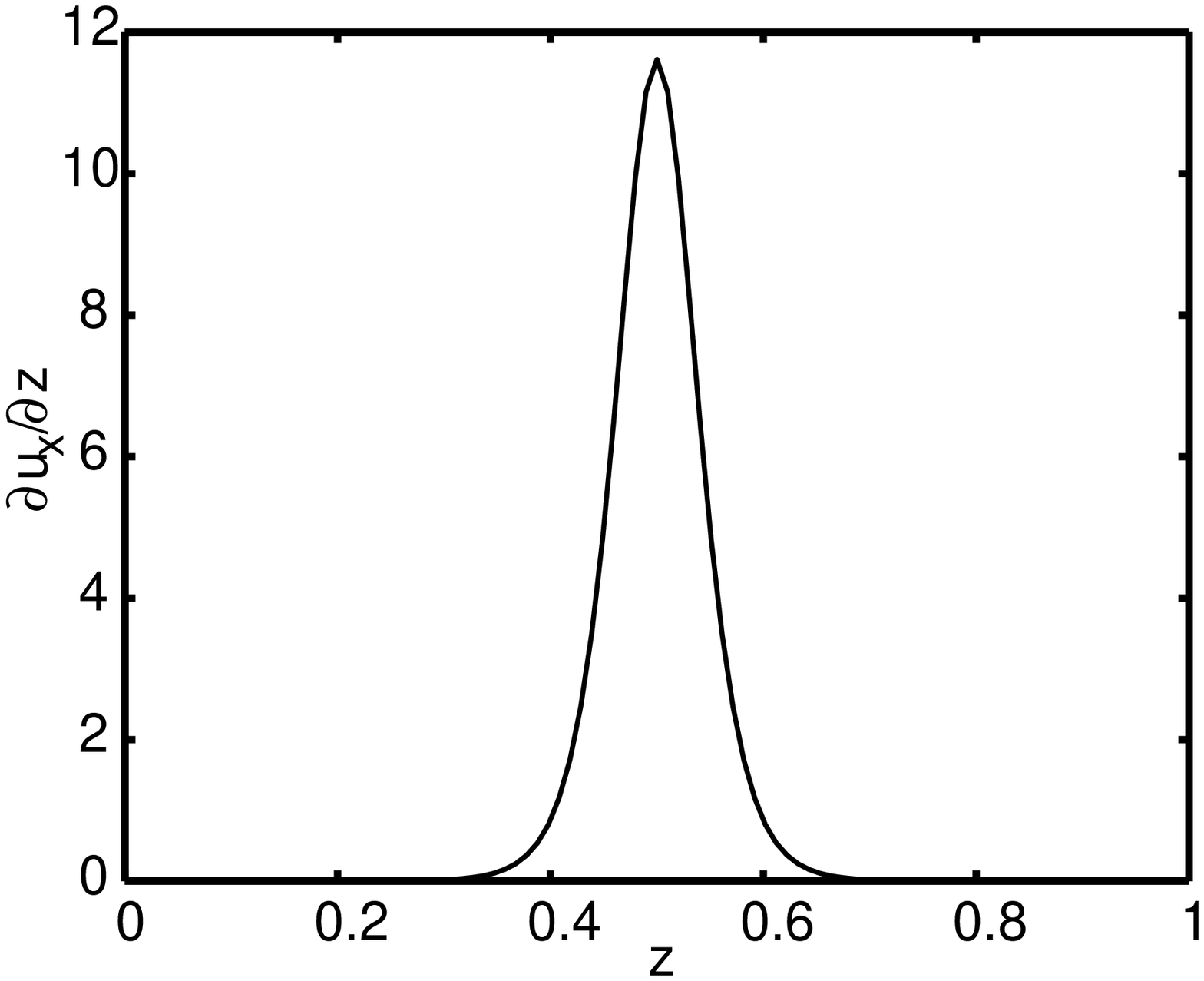,width=8cm, height=5cm}
\end{center}
  \caption{Top: $u_x$ as a function of depth for the single layer
  model. Bottom: $\partial{u_x}/\partial{z}$ as a function of depth for the single layer model. }\label{fig1}
\end{figure}

This system of equations has a large number of dimensionless parameters,
making it impractical to conduct a complete survey of parameter space.
Therefore we focus primarily upon varying the strength of the magnetic
field, holding all other parameters fixed (although a small number of
runs with different parameter values were also carried out). The
parameter choices are summarised in Table~\ref{table1}. Note that this
choice of $\theta$ implies that the layer is weakly
stratified. Setting $m=1.6$ and $\gamma=5/3$ implies that
stratification in the layer is mildly subadiabatic. This choice of
parameters is appropriate for the stably stratified solar tachocline
\citep{Vasila,CT}. Note also that the parameter values that are given
in Table~\ref{table1} imply that the dynamical effects of the magnetic
diffusivity, the viscosity and the thermal diffusivity are much more
significant in this model than in the solar interior. This is because
the dissipative length scales associated with the solar interior can
not be resolved using any current computer. However, by setting
$1>\zeta_0>\sigma$, we ensure that these are ordered in the same way
as in the solar interior, i.e. the thermal dissipative cutoff scale is
larger than the magnetic dissipative cutoff scale, which is in turn
taken to be larger than the viscous scale.

\par Finally, we must
specify a suitable initial shear flow for this system. We set

\begin{equation}
U_0(z)=0.577(1+ \textrm{tanh} [20(z-0.5)])\label{equation10}
\end{equation}

\noindent as shown in Figure~\ref{fig1}(top). The hyperbolic tangent
gives a smooth velocity field, varying from $u_x=0$ at $z=0$ up to
$u_x=1.154$ at $z=1$. The width of the shear region is sufficiently small
 that 
 the departure from a stress-free
condition at the boundaries is comparable with the numerical error of
the scheme. The
results of \citet{Vasila} suggest that a stronger shear promotes
magnetic buoyancy instabilities. We have
maximised the shear velocity subject to the constraint that the horizontal
flow speed never exceeds the adiabatic sound speed, whilst also
ensuring that the peak mach number of the flow is
identical to the peak mach number of the shear in the composite model
(see the next section). Note that the
fluid Reynolds number of this shear (based upon the peak velocity and
the width of the shear layer) is approximately $300$, which is much
smaller than in other studies \citep{Vasila,Vasilb,siletal09}. As
discussed in the Introduction, this enables us to carry out fully
resolved simulations with comparatively modest numerical grids. 

\subsection{Results}

\begin{figure}
\begin{center}
  \epsfig{file=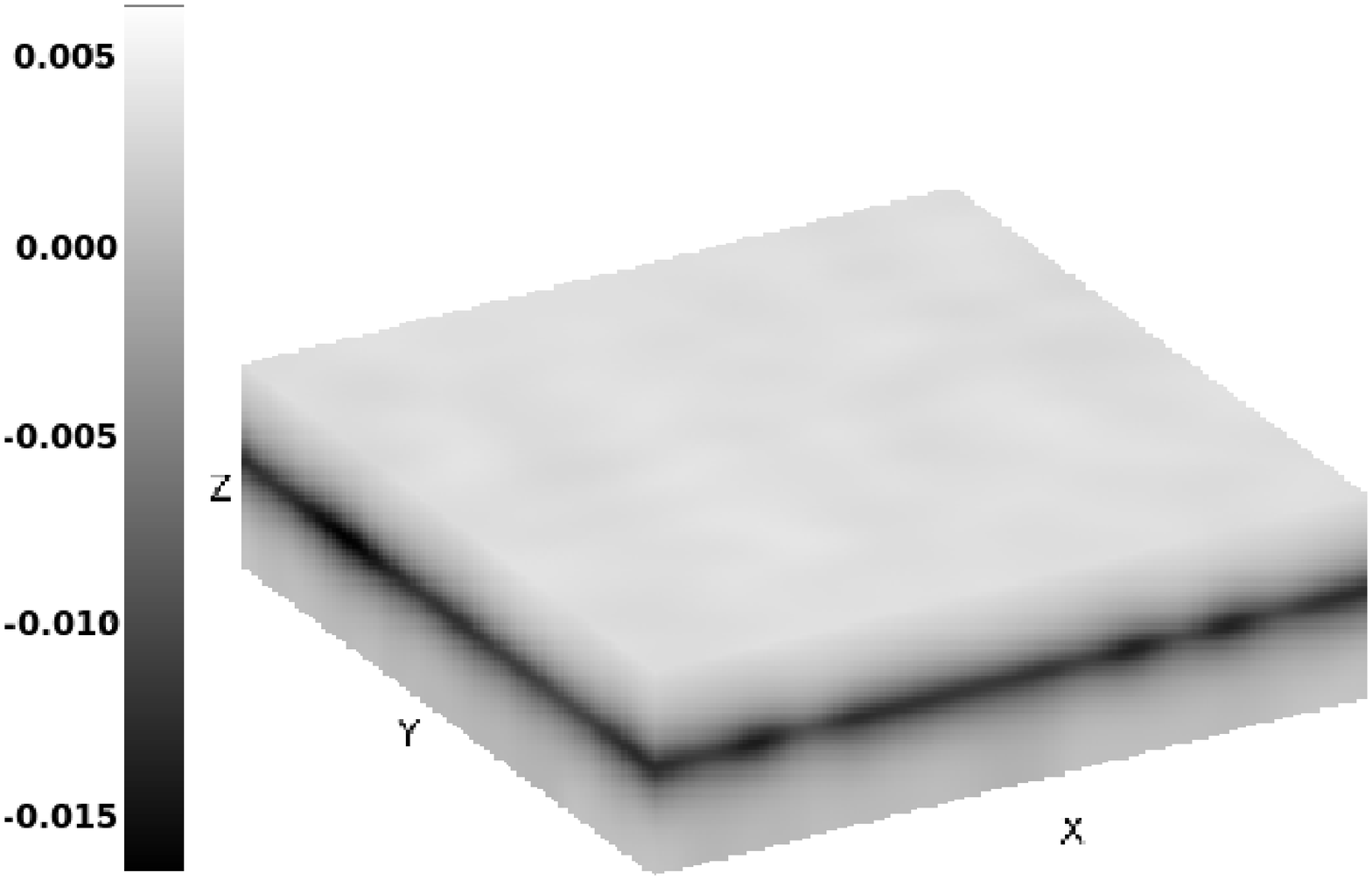,width=9cm, height=7cm}
\epsfig{file=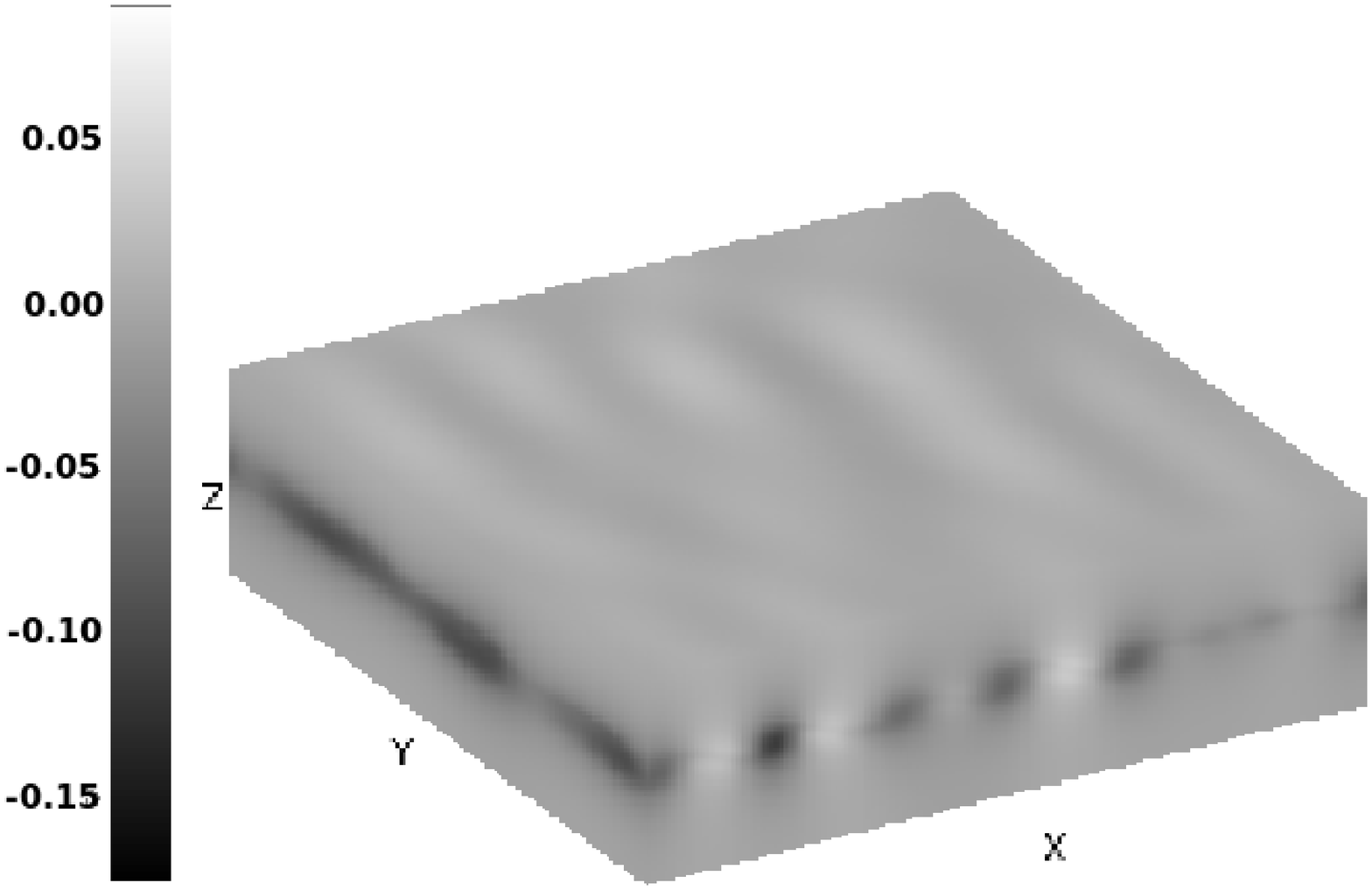,width=9cm, height=7cm}
\epsfig{file=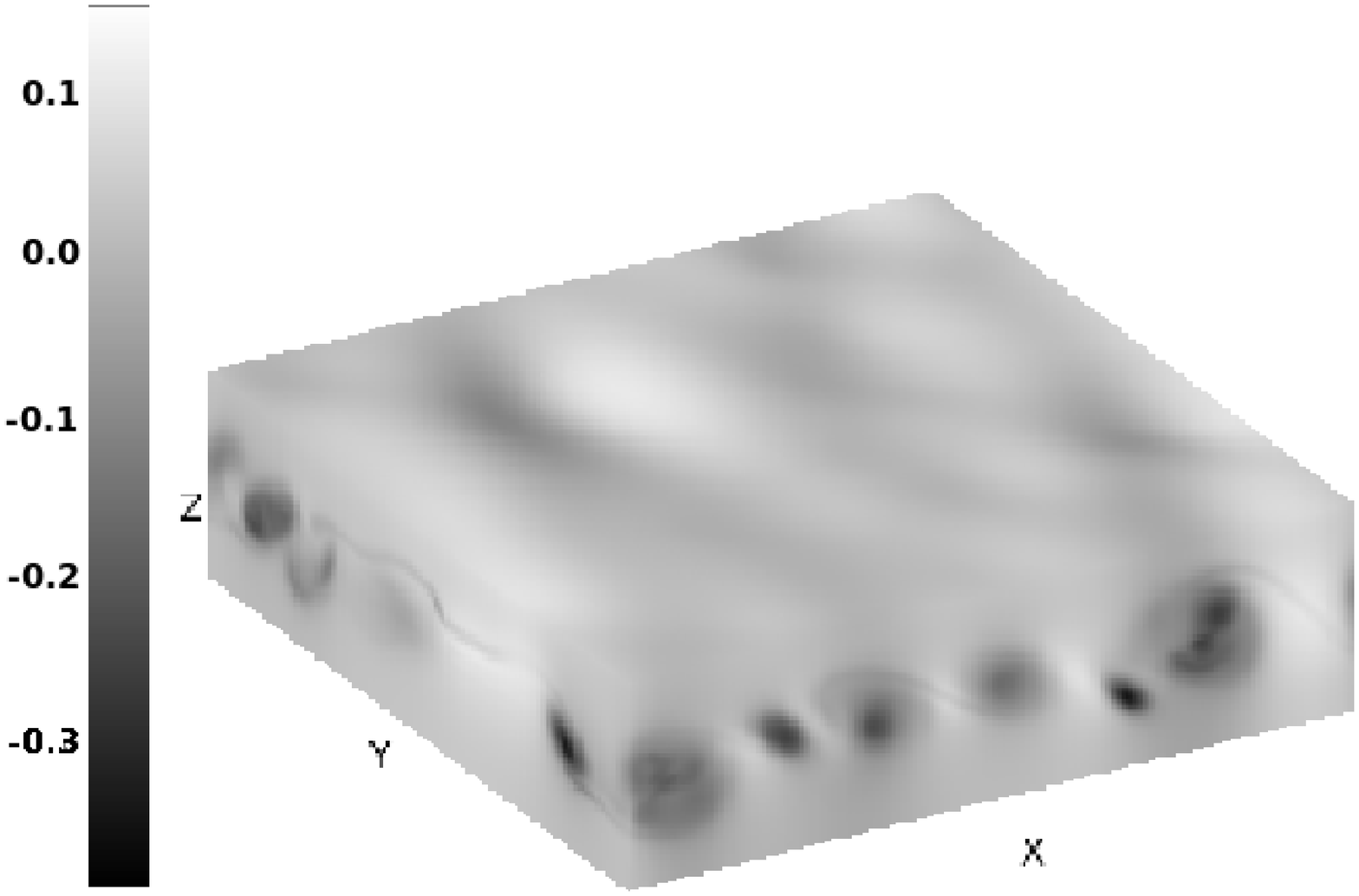,width=9cm, height=7cm}
\end{center}
  \caption{Density perturbation snapshot for the hydrodynamic case
    ($F=0$) at
  times $t=2.59$ (top), $t=5.20$ (middle) and $t=7.69$ (bottom)}\label{fig2}
\end{figure}

\begin{figure}
\begin{center}
\epsfig{file=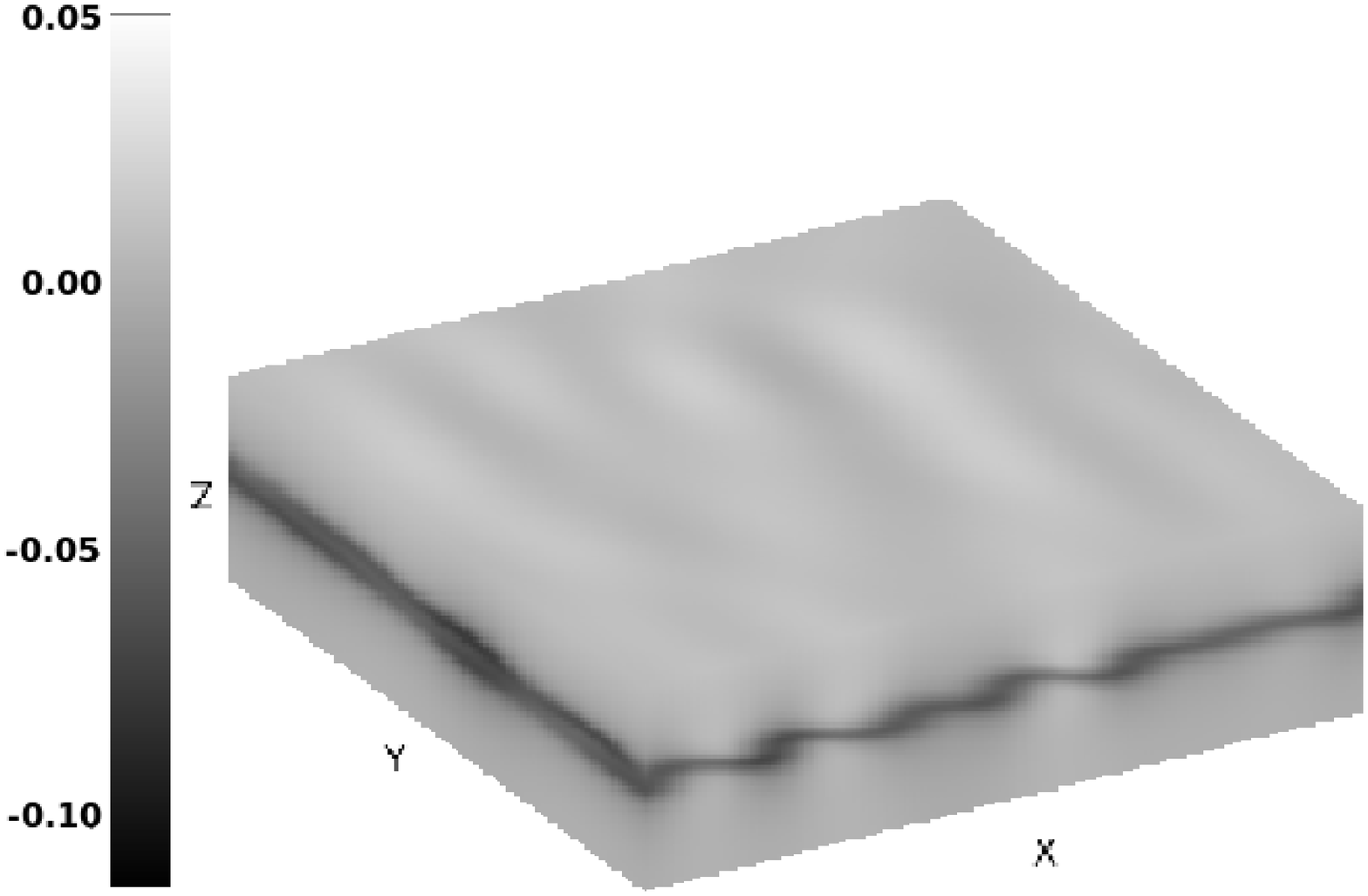,width=8cm, height=7cm}
\epsfig{file=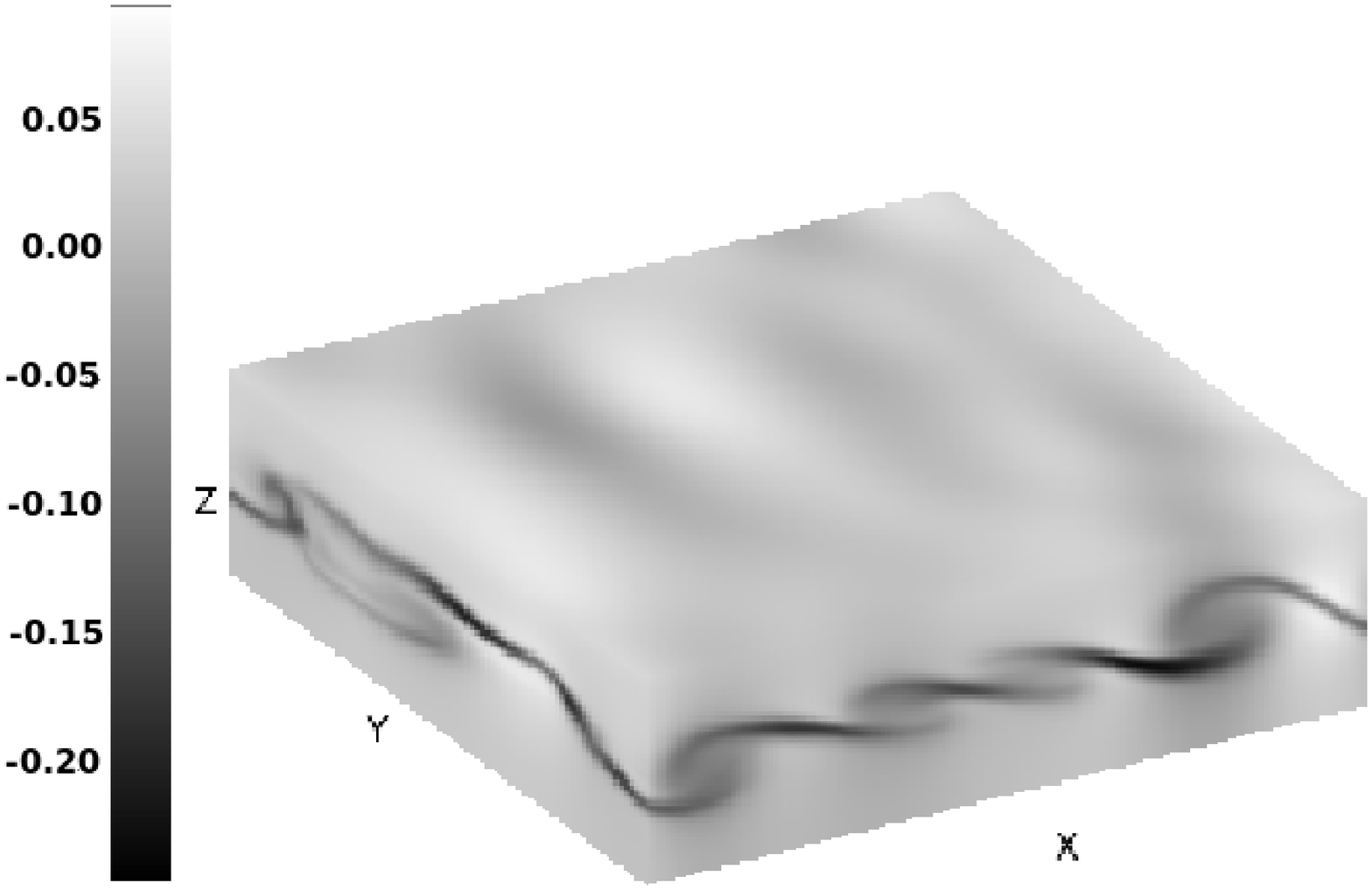,width=8cm, height=7cm}
\end{center}
  \caption{Density perturbation snapshot for F=1/9000, at t=5.21 and
  7.78.}\label{fig3}
\end{figure}

Having set up the model problem for a stably-stratified polytropic
layer, we now investigate the effects of varying the imposed
magnetic field. In order to achieve this, we carry out a series of
numerical simulations for different values of the parameter, $F$
(keeping all other parameters fixed, as shown in
Table~\ref{table1}). In the absence of a magnetic field ($F=0$),
we find that the system is unstable to a  shear flow (Kelvin-Helmholtz type)
instability. Initially, we see rolls forming in the $x-z$ plane, as shown in Figure~\ref{fig2}. This then rapidly
evolves into a three-dimensional time-dependent flow. The implications of
imposing a magnetic field across this computational domain depend
upon the strength of the imposed magnetic field. If the
initial field is weak (say $F=1/90000$), then the effect of the field
on the evolution of the instability is negligible, as the solutions
are virtually indistinguishable from the hydrodynamic case. The
magnetic field is simply advected with the resultant flow as a passive
vector field. However, if we increase the strength of the magnetic
field, we find that it starts to have a dynamical
influence. Figure~\ref{fig3} shows the evolution of a simulation with a
magnetic field strength determined by $F=1/9000$. In this case,
the magnetic field does reduce the vigour of the
instability, leading to more ordered motions (particularly at early
times). This behaviour is easy to explain. The shear flow instability
acts so as to bend the magnetic field lines parallel to the velocity
shear. A strong field tends to resist this process, thus inhibiting
the instability. However even in this simulation, as in the
hydrodynamic case, the instability eventually develops
three-dimensional structure. 

\begin{figure}
\begin{center}
\epsfig{file=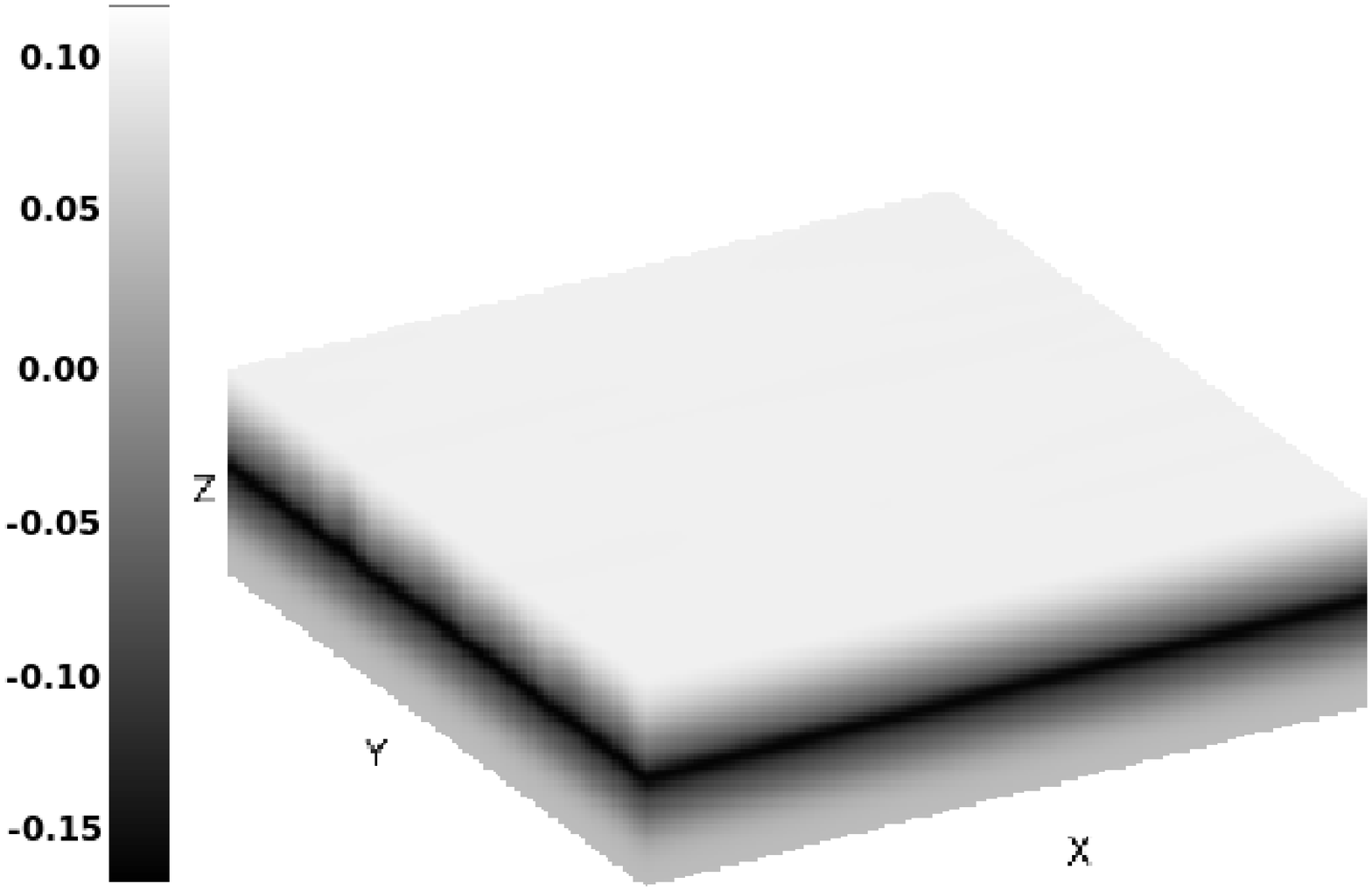,width=8cm, height=7cm}
\epsfig{file=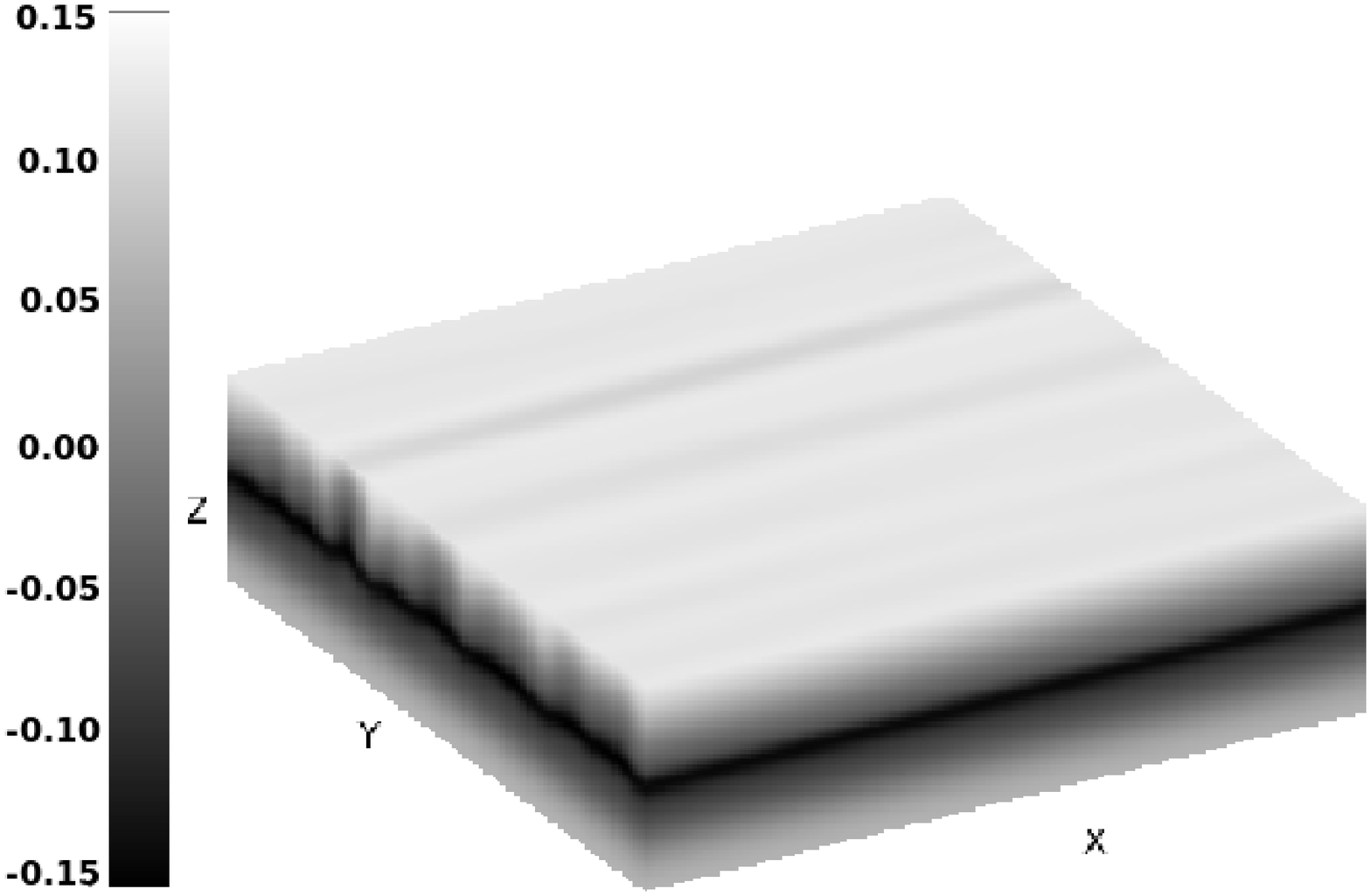,width=8cm, height=7cm}
\end{center}
  \caption{Density perturbation snapshot for F=1/900, at t=10.93 and
  13.88.}\label{fig4}
\end{figure}

The character of the instability changes further as we increase
the strength of the imposed magnetic field. Results for $F=1/900$ are
shown in Figure~\ref{fig4}. The resulting horizontal magnetic field is
now strong enough to completely suppress the shear flow
instability. Rather than generating fluctuations parallel
to the shear, the initial instability is a short wavelength
interchange instability, with almost all variation
(at least initially) in the $y$ direction. These interchange modes are
typical of a magnetic buoyancy instability
\citep{Newcomb1961,Hughes1985}. During these early stages, the
developing structures are similar to those found in two-dimensional
calculations of the break up of a magnetic layer in the absence of a
shear \citep{Catthughes}. At later times some longer wavelength
variation in the $x$ direction does appear -- this three-dimensional
evolution is similar to that found by \citet{Wissink}.

While the focus of this section is to explore the effects of varying
the magnetic field strength, we note that there are other parameters
that can be varied (subject to computational constraints). A reduction
in the Prandtl number leads to a reduction in the viscous dissipation
relative to the other diffusivities. This also increases the fluid
Reynolds number. We  carried out runs with lower values of the Prandtl number, and found that reducing this parameter  by up to a factor of ten (going down to  $\sigma\sim 0.005$) appears to
have little effect upon the vigour of the instability at
$F=1/900$. This suggests that there is very little dependence upon the fluid
Reynolds number in this parameter regime. The effects of
varying both $\kappa$ and $\sigma$ have not been systematically
studied here;  \citet{siletal09} have shown
that $\kappa$ in particular  can play an important role for larger values of the Richardson number, and investigation of a full range of diffusivity ratios will be the subject of future work.

\section{The Composite Model}

In this section, we consider a more complicated model problem,
consisting of a piecewise polytropic atmosphere. This is intended to
be a highly idealised representation of the region straddling the base
of the solar convection zone. In order to achieve this, we consider a
deeper computational domain, corresponding to $z_0=3$. We also choose
a wider computational domain, by setting $\lambda=8$. Recalling the
definitions of these parameters, this implies that the computational
domain is defined by $0\le x,y\le 8$ and $0\le z\le 3$.

\subsection{Parameters}

Other than the dimensions of the computational domain, the main difference
between these calculations and those of the preceding section is
that the polytropic index of the domain is now a function of
depth. We split up the domain into
three layers of unit depth. In the top layer ($0 \le z \le 1$), we
choose a polytropic index of $m_0=1$, which implies that this region
is convectively unstable. Like the single layer from the previous
section, the middle region ($1 \le z \le 2$) is convectively stable
with $m_1=1.6$. The lower layer ($2 \le z \le 3$) is also convectively
stable but with a much larger polytropic index, $m_2=4$. The primary
purpose of the lower layer is to lessen the impact of the rigid lower
boundary. Any descending convective plumes that reach $z=2$ can simply
pass into the lower region without ``splashing'' back and interfering
with the other dynamics in the system. In order to achieve the
required piecewise-polytropic structure, we choose the following
depth-dependent thermal conductivity profile: 

\begin{eqnarray}
\label{equation11}
K(z) & = &\frac{K_0}{2}\left[1-\tanh\left(\frac{z-1}{0.02}\right)\right]\\
\nonumber  &+&\frac{K_0\left(m_2+1\right)}{2\left(m_0+1\right)}\left[1+\tanh
\left(\frac{z-2}{0.02}\right)\right]\\ \nonumber
&+&\frac{K_0\left(m_1+1\right)}{2\left(m_0+1\right)}\left[1
  -\tanh \left(\frac{z-2}{0.02}\right) \tanh\left( \frac{z-1}{0.02}\right) \right],
\end{eqnarray}

\noindent where $K_0=K(0)$ (as before). The $\tanh$ profiles ensure
that the conductivity varies smoothly between each region. 
\begin{figure}
\begin{center}
\epsfig{file=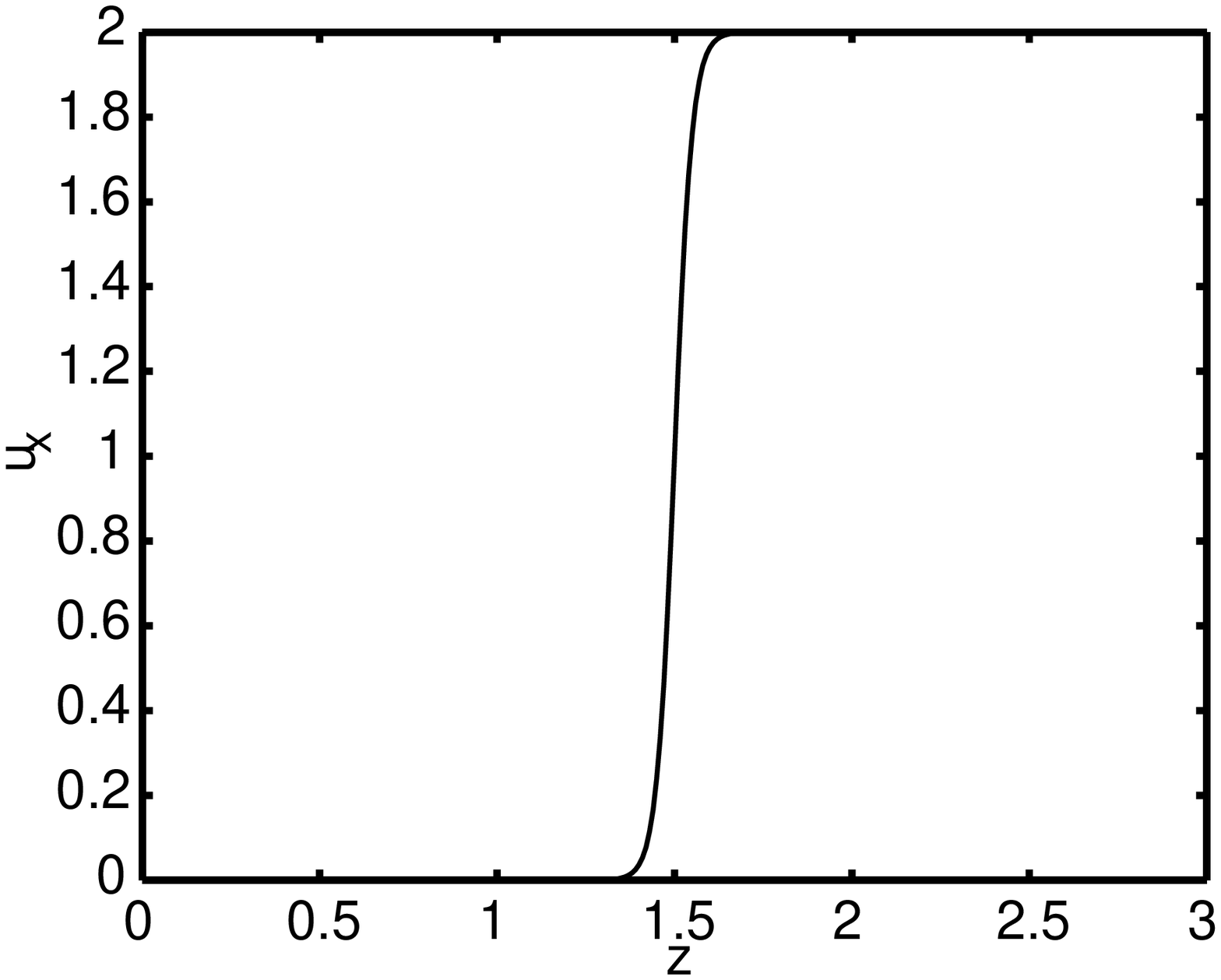,width=8cm, height=5cm}
\epsfig{file=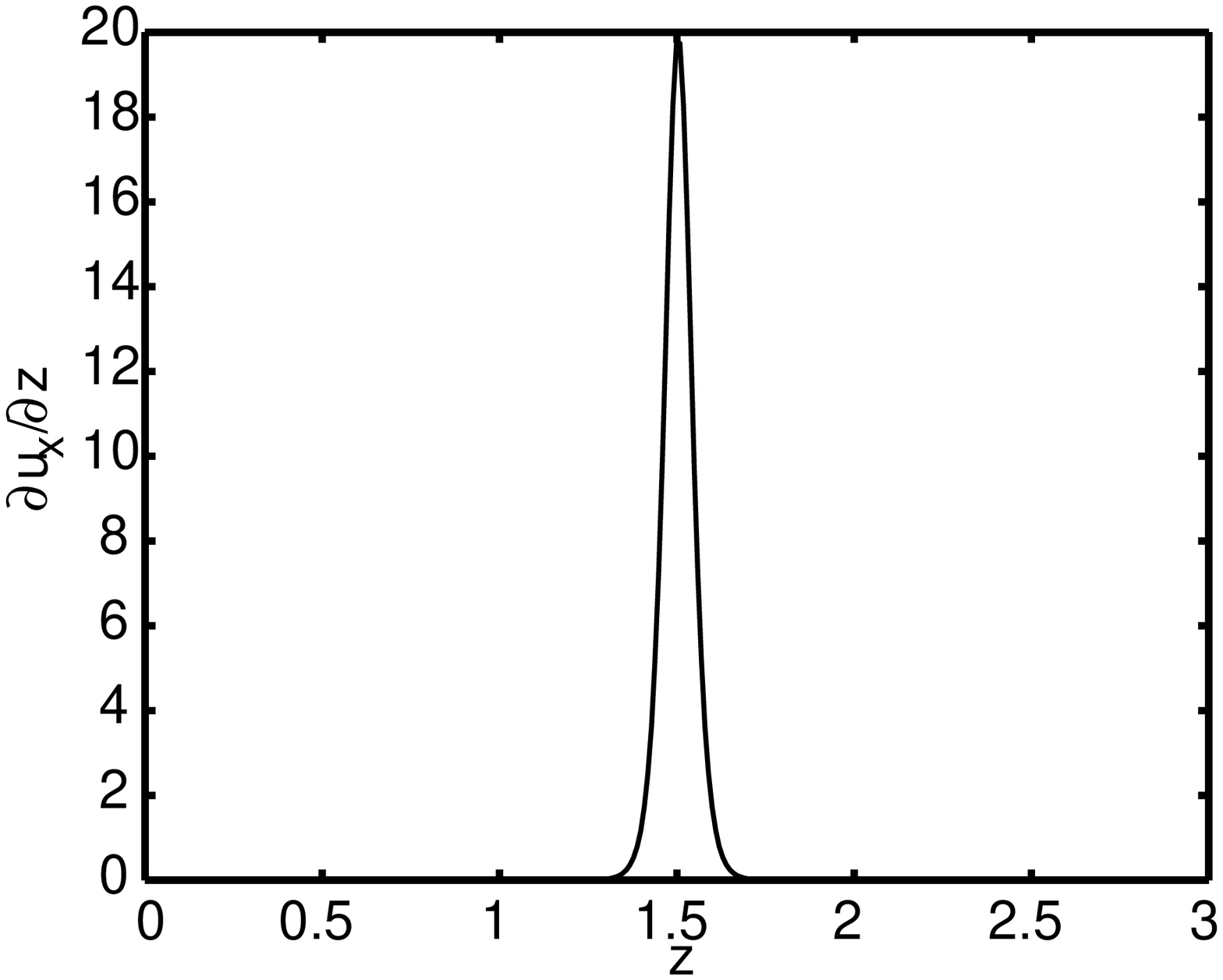,width=8cm, height=5cm}
\end{center}
  \caption{Top: $u_x$ as a function of depth for the composite layer
  model. Bottom: $\partial{u_x}/\partial{z}$ as a function of depth for the composite layer model.}\label{fig5}
\end{figure}
\par In this composite model, our aim is to investigate the
effects that any shear instabilities in the mid-layer have upon an
established pattern of convection. Therefore, we only introduce the
shear once the convection in the upper layer has become fully
developed. This is achieved by integrating the equations without any
horizontal forcing until $t\approx40$. The shear is then introduced at
this point, along with the corresponding forcing term in
Equation~\ref{equation2}. Once it has been introduced, the shear has
the same structure as that in the single layer model but is now
centred at the mid-plane of the middle region (at
$z=1.5$). Thus the imposed shear now has the form:

\begin{equation}
U_0(z)=1+ \textrm{tanh} [20(z-1.5)],\label{equation12}
\end{equation}

\noindent as shown in figure~\ref{fig5}. Note that the amplitude of
the shear is chosen so that the local Mach number of the flow is the
same as for the single-layer case.

\par
\begin{figure}
\begin{center}
\epsfig{file=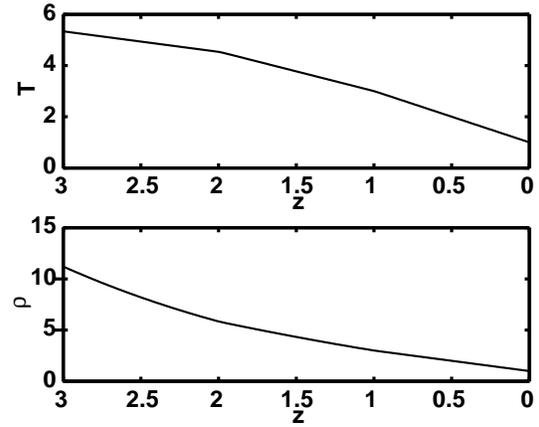,width=8cm, height=6cm}
\end{center}
  \caption{The initial temperature and density profiles for the composite model.} \label{fig6}
\end{figure}

 In the absence of any imposed shear at $t=0$, the initial
conditions for this model differ slightly from those in the single
layer case. Here we choose a magnetohydrostatic initial condition, setting
$B_z=1$ and $u_x=u_y=u_z=B_x=B_y=0$. The equilibrium profiles for $\rho(z)$ and
$T(z)$ are found numerically, and are shown in Figure~\ref{fig6}. Note
that the choice of the thermal boundary condition at the lower surface
determines the extent of the thermal stratification. Setting $\partial
T/\partial z = 0.8$ at $z=3$ ensures that the temperature increases
by $50\%$ across the middle layer, as was the case for the single layer.

\par The parameters for this composite model are chosen
so that the conditions in the middle layer are as similar as possible
to those for the single layer calculation. Note that this requires
some rescaling of $\kappa$ and $\zeta_0$. These parameters are shown
in Table \ref{table2}.

\begin{table}
\begin{center}
\begin{tabular}{|c|c|c|}
\hline
Param.   &   Description &  Value  \\
\hline
$\sigma$  &  Prandtl  Number  &  0.05 \\
$m_0, m_1, m_2$   &  Polytropic  Indices &  1.0, 1.6, 4.0  \\
$\gamma$  &  Ratio  of  Specific  Heats & $5/3$ \\
$\kappa$  &  Thermal Diffusivity & 0.0385   \\
$F$ & Magnetic Field Strength& variable \\
$\zeta_0$ & Magnetic Diffusivity & 0.1\\
\hline
\end{tabular}
\caption{Fixed Parameter Values} \label{table2}
\end{center}
\end{table}

\subsection{Results}

\begin{figure}
\begin{center}
\epsfig{file=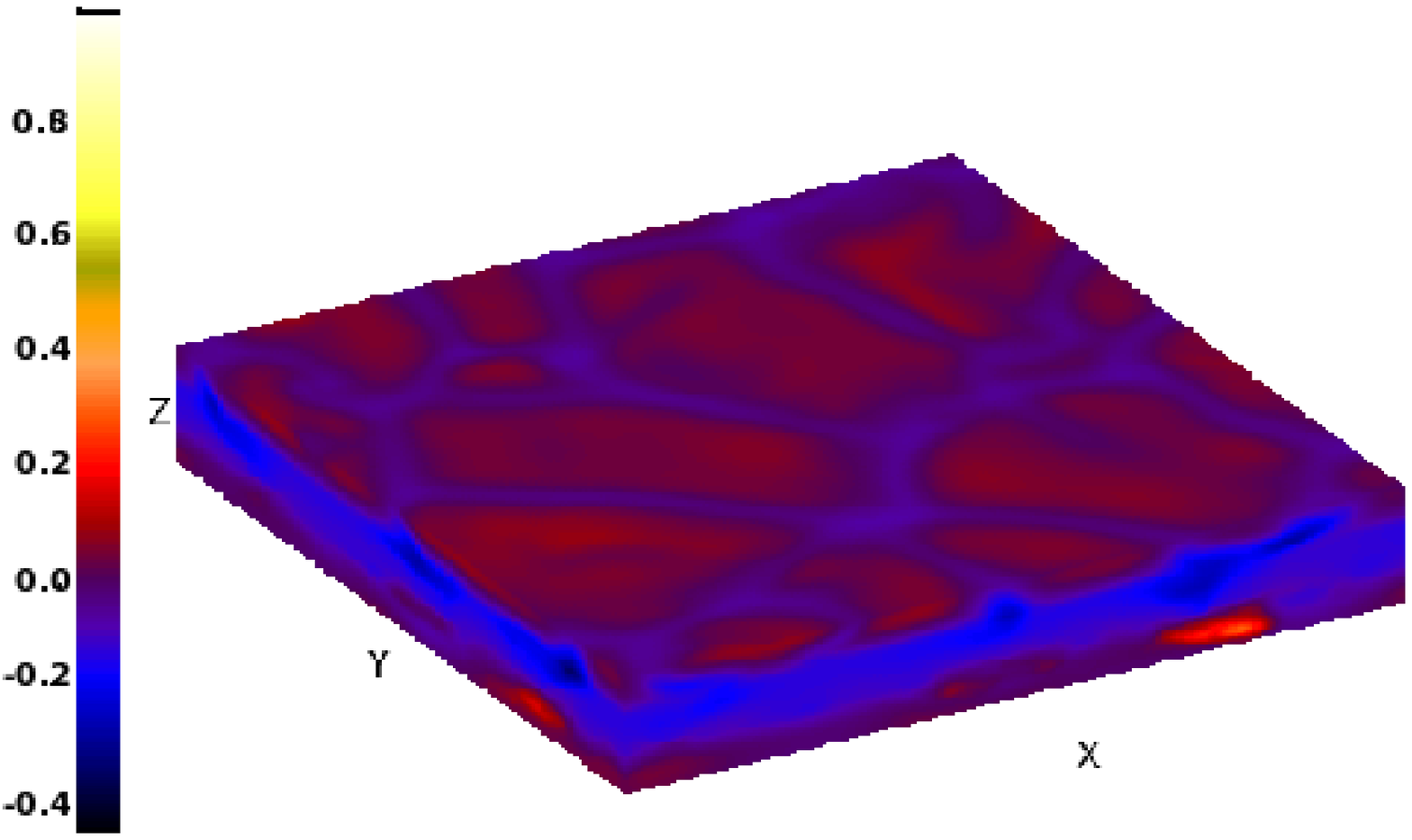,width=9cm, height=9cm}
\epsfig{file=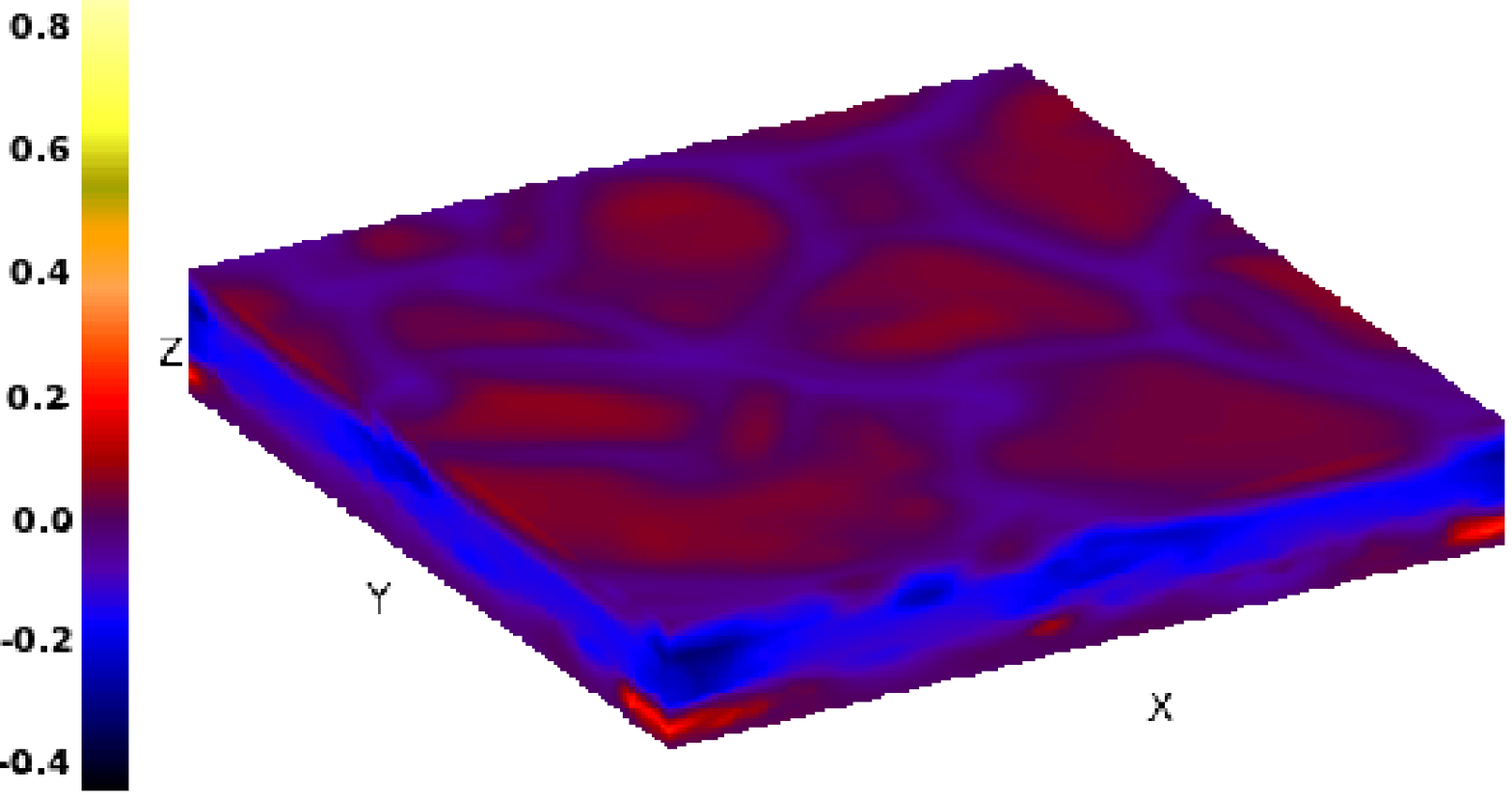,width=9cm, height=9cm}
\end{center}
  \caption{Three-dimensional plots of the $F=0.0001$ case in the
    absence of any velocity shear. Top: The temperature perturbation
    on the side of the computational domain and close to the upper
    surface at $t\approx40$. Bottom: The same plot at $t\approx51$. }
  \label{fig7}
\end{figure}
\begin{figure}
\begin{center}
\epsfig{file=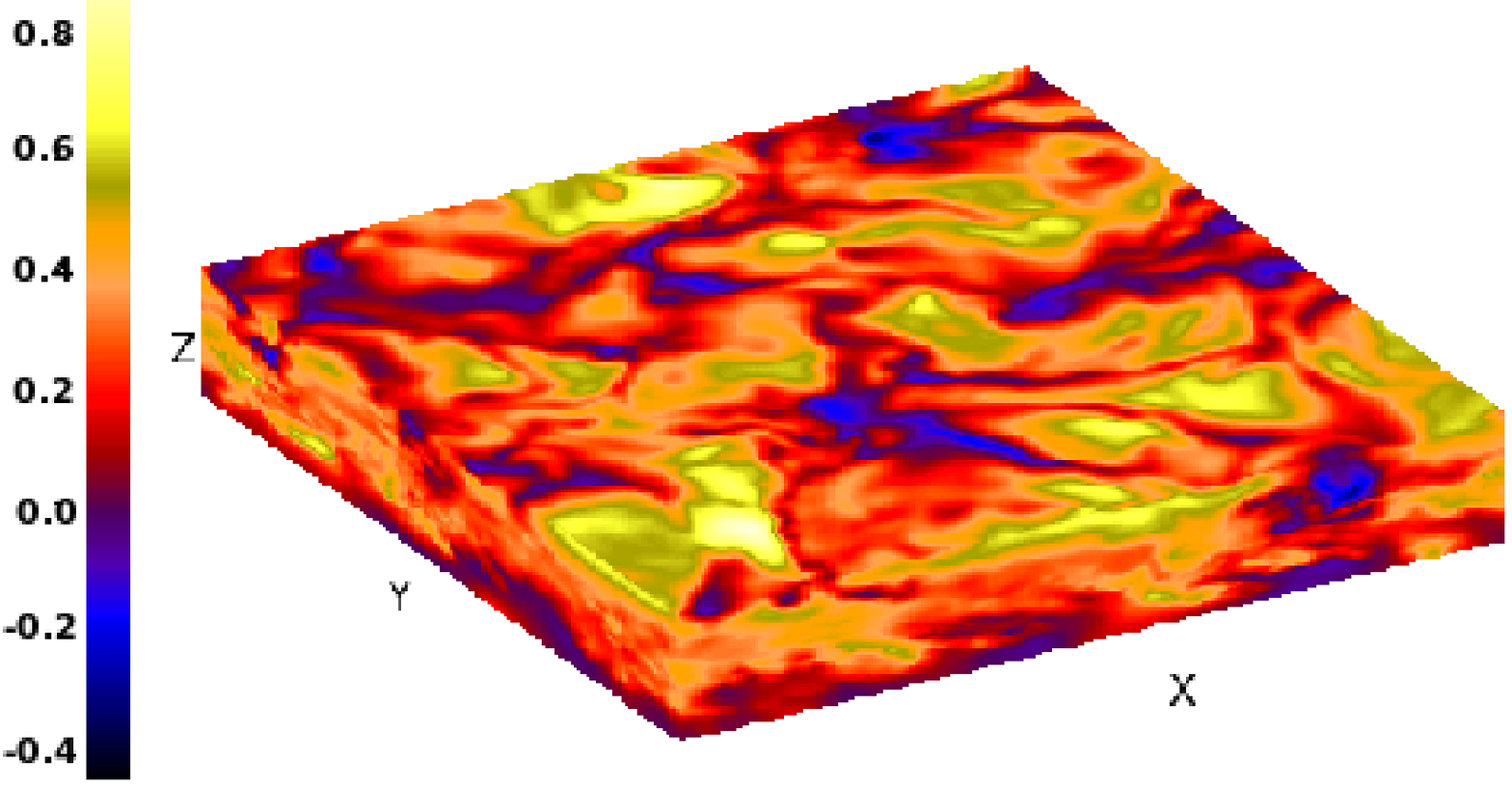,width=9cm, height=9cm}
\epsfig{file=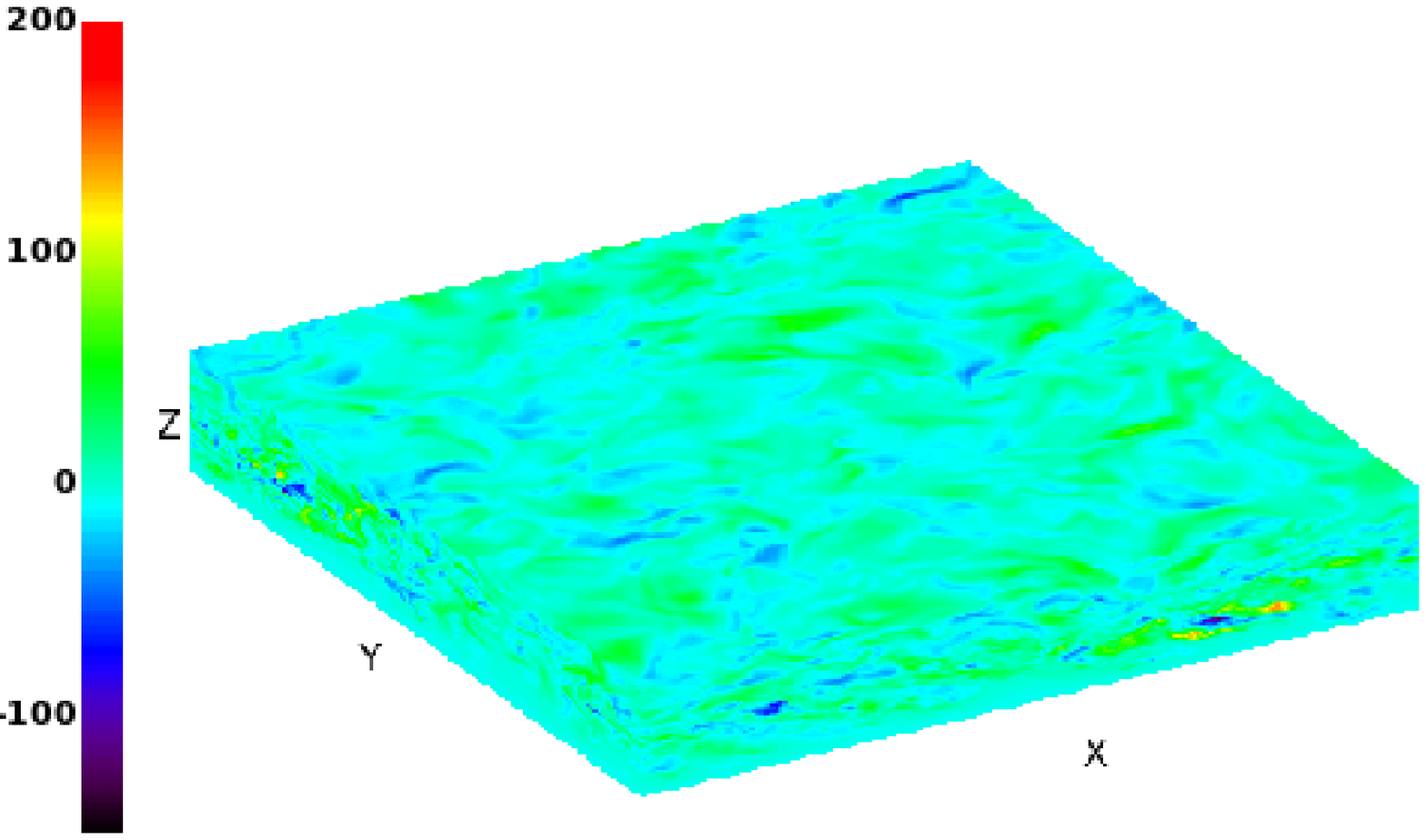,width=9cm, height=9cm}
\end{center}
  \caption{Three-dimensional plots of the $F=0.0001$ case at
    $t\approx51$, after the shear was introduced at $t\approx
    40$. Top: The temperature perturbation on the side of the
    computational domain and close to the upper surface. Bottom: The
    horizontal magnetic field on the same surfaces. Note that in this
    figure, $B_x$ is normalised with respect to the imposed field,
    $\propto \sqrt{F}$.}
  \label{fig8}
\end{figure}

As for the single layer calculations in the previous
section, we carry out a series of numerical simulations for
different values of the imposed magnetic field (as measured by
$F$). In addition to any effects upon the hydrodynamic instabilities
of the shear, increasing $F$ also reduces the vigour of any convective
motions in the upper region of the computational domain.\footnote{Note
  that estimates from linear theory suggest that a value of F of
  approximately 0.5  is needed in order to completely suppress
  convective flows in the upper layer. Therefore, we are not near the
  convective stability boundary. If we were to increase F further, we would expect to see a transition to an oscillatory mode of convection before we reach the regime in which convection is completely inhibited. } The range of
$F$ is carefully chosen so that we cover the same values for
the mid-layer plasma beta (the ratio of the gas pressure to the
magnetic pressure) that were covered in the single-layer case. 

Initially, we set $F=0.0001$, which corresponds to a weak imposed
magnetic field. Given the relative complexity of this system, we first
explore the dynamics that occur in the absence of any velocity
shear. This case is illustrated in Figure~\ref{fig7}, which shows the
resulting pattern of convection at $t\approx40$ (top) and $t\approx
51$ (bottom). In the convectively unstable upper layer there is a
time-dependent cellular convective pattern consisting of warm, broad
upflows (which correspond to the brighter regions in
figure~\ref{fig7}) surrounded by a network of narrower (darker)
downflows. There is some modest convective overshooting into the middle
layer. Figure~\ref{fig8} also shows a snapshot of this system at
($t\approx51$) but, in this case, the shear is introduced at $t\approx
40$. As was found in the
single-layer case, the shear is subject to a Kelvin-Helmholtz type
shear instability. This rapidly develops three-dimensional structure,
producing perturbations that spread throughout the stably-stratified
domain, penetrating into the convective layer. Comparing Figure~\ref{fig7}
and \ref{fig8} we see that there is little evidence of anisotropy in
the convecting region. However  we note that there is a significant influence on the convective
transport of heat in the top part of the box, which is apparently due to the influence of the shear instability on heat transport in the middle layer.

\begin{figure}
\begin{center}
\epsfig{file=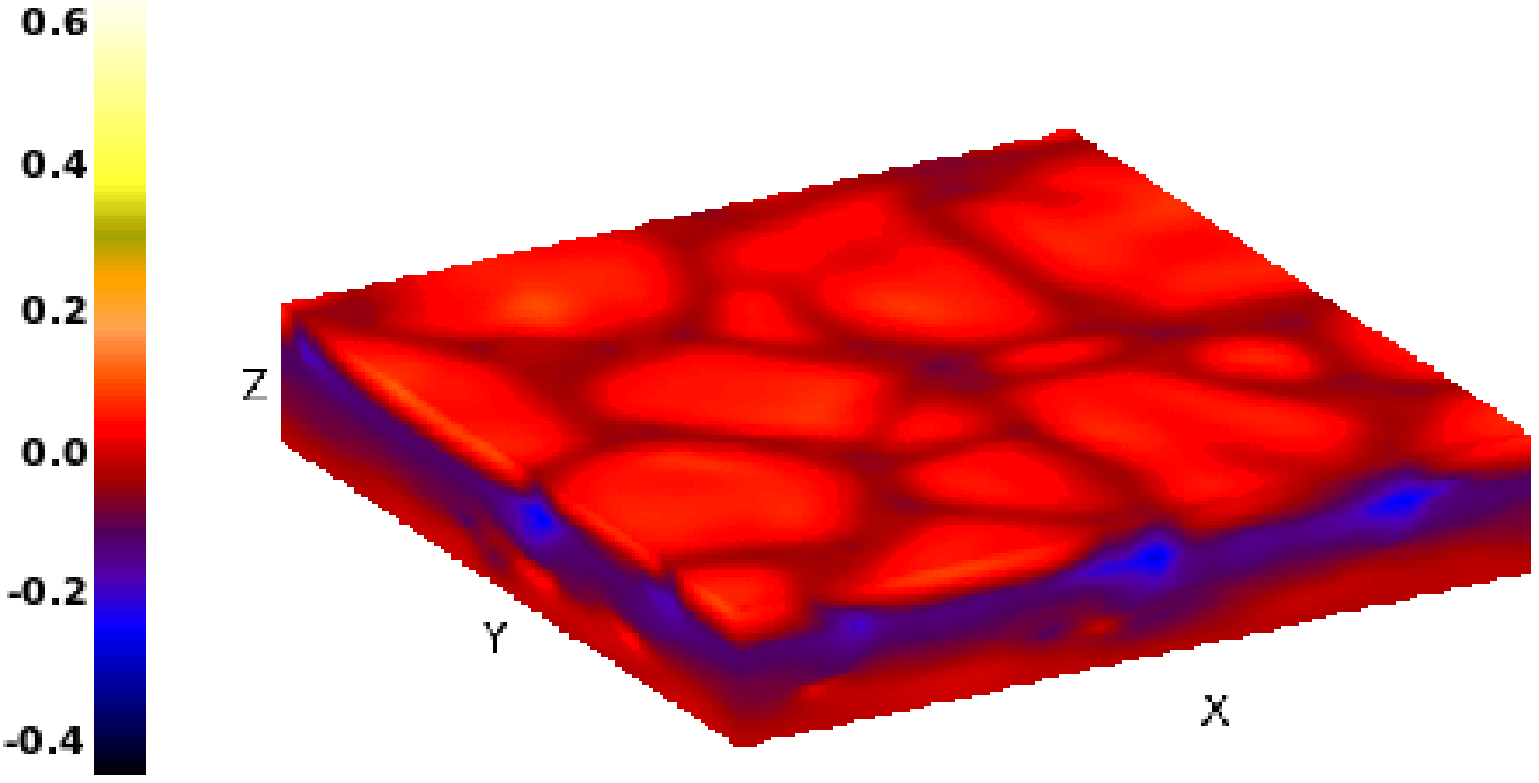,width=9cm, height=9cm}
\epsfig{file=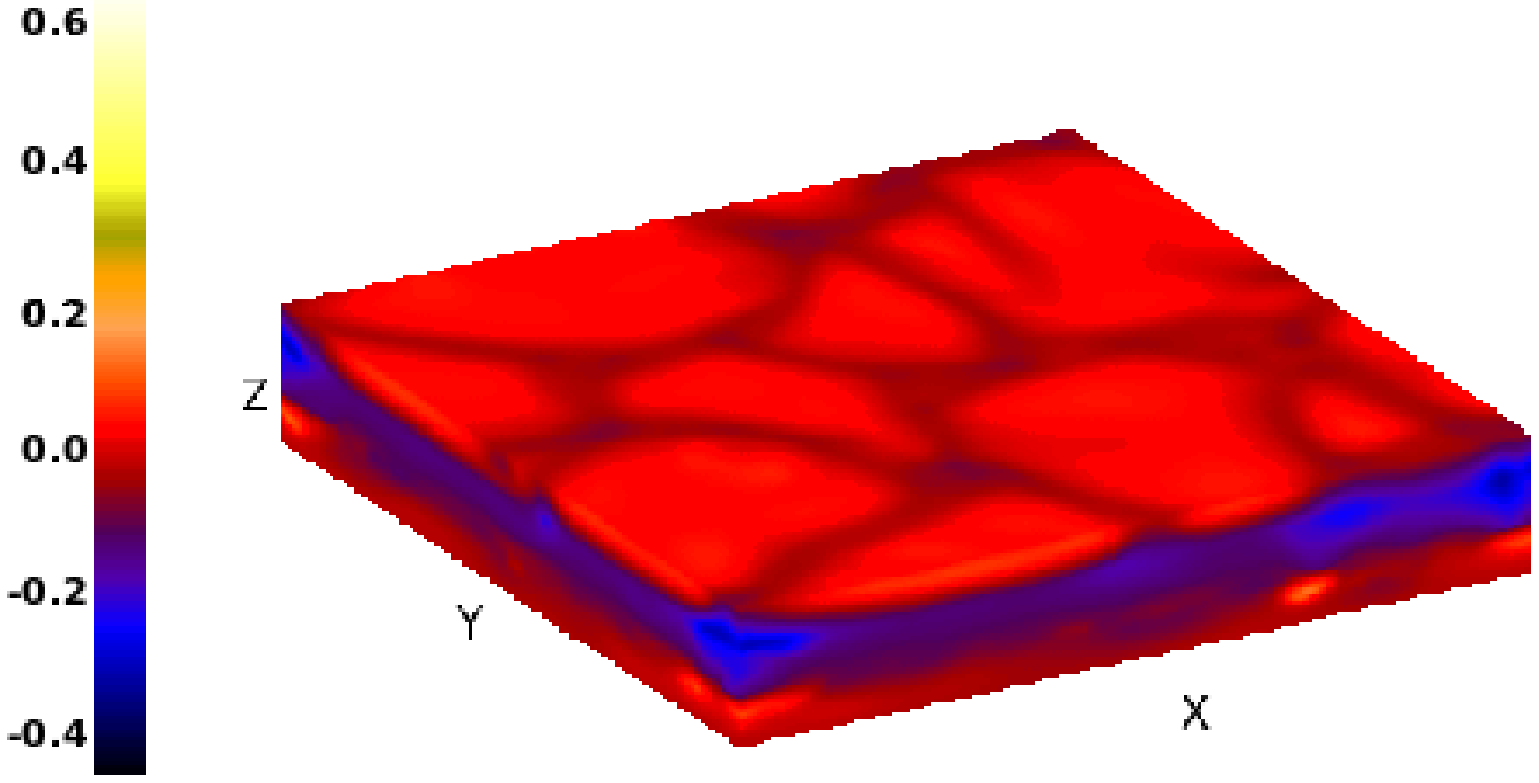,width=9cm, height=9cm}
\end{center}
  \caption{Three-dimensional plots of the $F=0.01$ case in the
    absence of any velocity shear. Top: The temperature perturbation
    on the side of the computational domain and close to the upper
    surface at $t\approx40$. Bottom: The same plot at $t\approx51$. }
 \label{fig9}
\end{figure}
\begin{figure}
\begin{center}
\epsfig{file=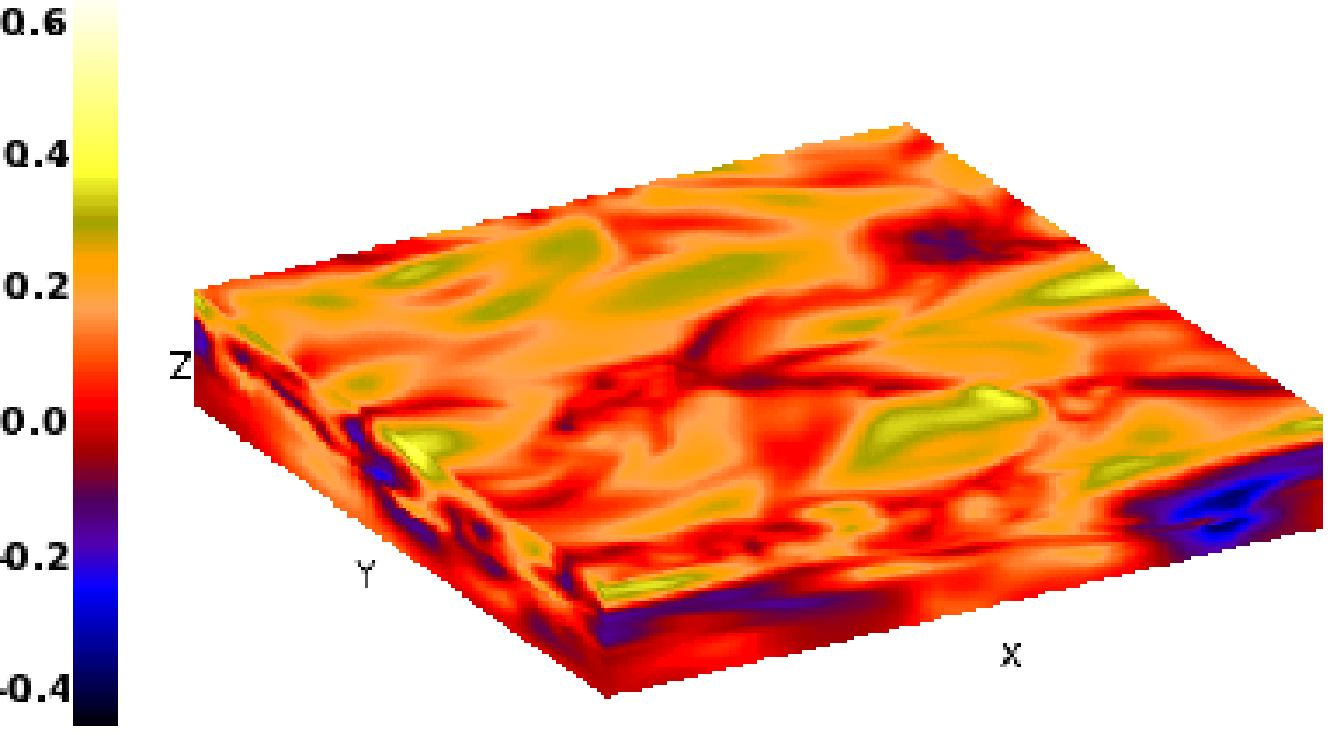,width=9cm, height=9cm}
\epsfig{file=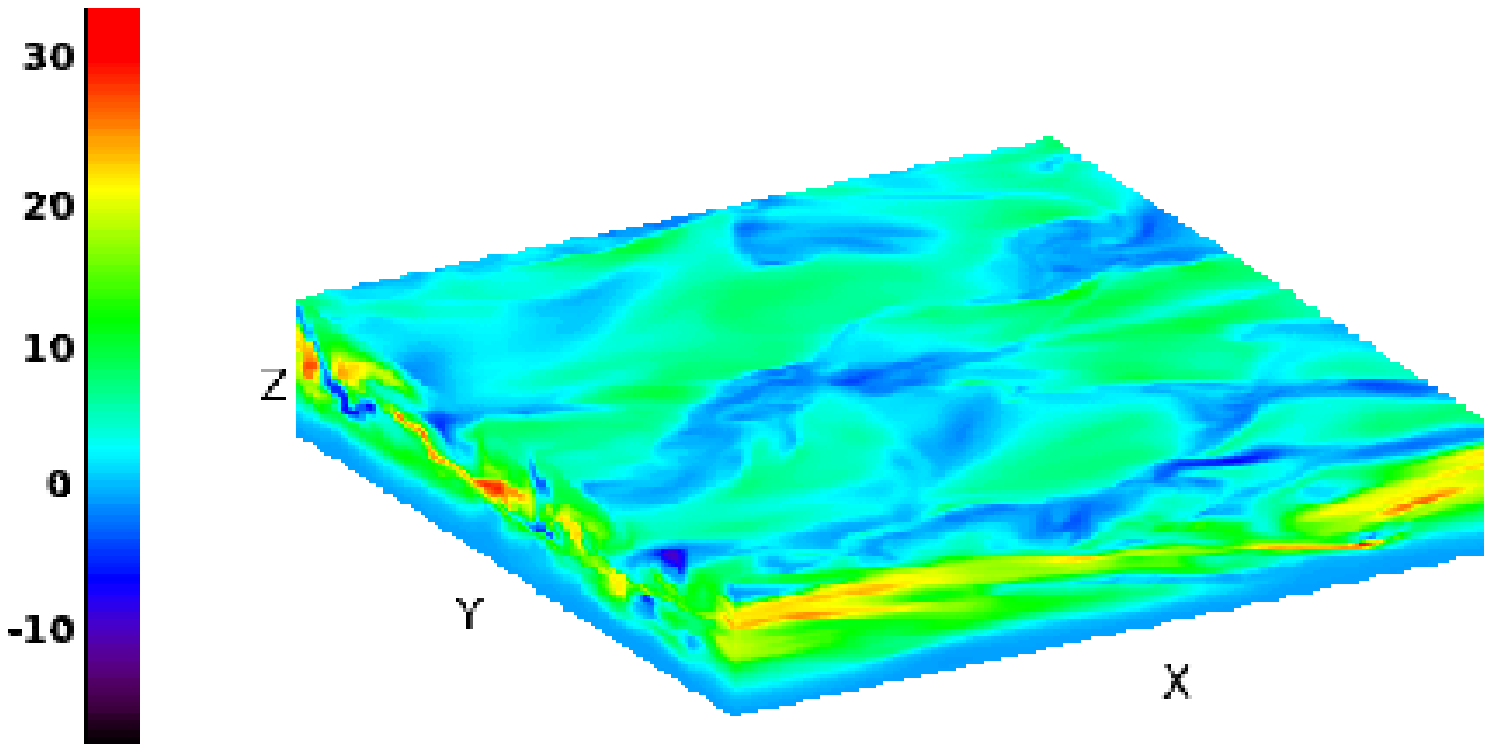,width=9cm, height=9cm}
\end{center}
  \caption{Three-dimensional plots of the $F=0.01$ case at
    $t\approx51$, after the shear was introduced at $t\approx
    40$. Top: The temperature perturbation on the side of the
    computational domain and close to the upper surface. Bottom: The
    horizontal magnetic field on the same surfaces. Note that in this
    figure, $B_x$ is normalised with respect to the imposed field,
    $\propto \sqrt{F}$.}
  \label{fig10}
\end{figure}

As the field strength is increased from this (effectively) kinematic
level, the solutions follow a similar trend to the single-layer
case. When $F=0.001$, the dynamics in the mid-layer are similar in
form to those shown in Figure~\ref{fig3} (although the overshooting
convection from the upper layer adds some additional complexity to the
resulting flows). Therefore, the dominant instability is still of
Kelvin-Helmholtz type rather than a magnetic buoyancy instability,
although magnetic effects are starting to play a dynamical
role. Interestingly, as the shear-driven motions from the stable layer
interact with the convective layer, there appears to be a slight
tendency for an elongation of the convective cells in the direction of
the shear. This is a phenomenon that becomes more pronounced as $F$ is
increased.

Increasing the field strength still further, so that $F=0.01$,
we find that the dynamics change dramatically. This case is
illustrated in Figures~\ref{fig9}
and~\ref{fig10}. The imposed vertical magnetic field is now strong
enough to reduce the vigour of the convection, though it has little
effect on the horizontal scales of motion. Once the shear is
introduced, the evolution is dominated by the shear-driven
instabilities at the mid-layer. As in the single layer case, a
transition has occurred so that the dominant instability is now
magnetic buoyancy. Initially, this buoyancy instability takes the form
of a two-dimensional (interchange) mode, although it soon develops
three-dimensional structure, forming arching regions of magnetic flux that rise up through the convective upper layer
of the domain. As these magnetic regions reach the upper layers, we
see some concentration of the vertical magnetic flux, which forms
localised concentrations near the horizontal boundaries of these
rising features. The subsequent motion is now strongly anisotropic, producing
convective cells that are predominantly aligned with the direction of
the shear and the buoyant horizontal magnetic flux
concentrations. We note that the introduction of a shear flow at
$t\approx 40$ again leads to larger temperature deviations in the
convectively-unstable region at later times (compared with the
unsheared case). However this effect seems to become less pronounced
as the field strength is increased. We attribute this phenomenon to the
fact that there is less mixing in the stronger field regime, where
there are larger structures present than in the weaker field cases. 

The phenomena discussed above can be related to previous work on
isolated buoyant flux tubes rising through the convection zone
\citep[see, for example,][]{Jouve}. Even though the tubes are isolated
they have been shown to interact strongly with the convective
flow. The present problem is different in that the magnetic field is
initially vertical. However, the shear creates a strong horizontal magnetic
field and so the ultimate configuration is not
dissimilar.

\section{Conclusions}

In this paper, we have presented some novel calculations to investigate
the ways in which an imposed magnetic field interacts with a shear
flow in a convectively-stable layer (both with and without an
overlying convective region). This investigation was motivated by
conditions at the base of the convection zone, and is relevant to the
interface scenario for the solar dynamo. The most important interactions between the shear layer and the convective region are due to the rising plumes
that are induced by magnetic buoyancy. In the solar context the tachocline
region, where the shear is expected to reside, is very stably
stratified and there are questions regarding the efficacy of magnetic
buoyancy in this situation \citep[see, for
example,][]{Vasilb}. Nonetheless we know that buoyancy instabilities
do occur in the Sun, and so it seems worthwhile trying to understand
aspects of their evolution, even though the correct parameter range
might not yet have been reached. In this context, it is also worth
noting that a diffusive instability, which is
effective only when the thermal diffusivity is much higher than in the
present paper, appears to allow for buoyancy-induced motion even when
the shear is hydrodynamically stable, according to the Richardson
number criterion. This mechanism been discussed for several decades
\citep[see, for example, the discussions in][]{Gilman,Hughes2007}  but
has only recently been demonstrated numerically in our geometry \citep{siletal09}.

In our ``strong-shear'' parameter regime, there is an instability of the shear (of Kelvin-Helmholtz type) even
when there is no magnetic field. This instability leads initially to
perturbations that are primarily in the $x-z$ plane. This
instability subsequently develops structure in the $y$ direction. We
find that, for a sufficiently strong magnetic field, the hydrodynamic
instability is suppressed, leading to a magnetic buoyancy instability
with strong variations in the $y$ direction, i.e.\ the direction
perpendicular to both gravity and the shear flow. For calculations of
the ``composite model'', the most important interactions between the
shear layer and the convective region are due to the rising plumes
that are induced by this magnetic buoyancy instability. This
instability generates strong horizontal concentrations of magnetic
flux that then rise through the convective layer. For the strongest
fields surveyed, the dynamical effects of these flux concentrations
are so significant that they destroy the convective pattern that would
normally exist in the absence of a magnetic field. As with isolated
flux tubes \citep[see, for example,][]{Jouve} they effectively push the
convective motion aside. It is important to note that even our
strongest imposed field has a relatively weak effect upon the
horizontal scales of convection that occur in the absence of shear.

Our results are encouraging in that they show that there exists the
possibility of inducing strong buoyant magnetic flux structures through the
action of horizontal shear. However they are preliminary calculations
in the sense that they do not allow the buoyant flux structures to
rise very far. We intend to perform further  calculations, with a much
deeper convective zone, so as to understand the later evolution of the
rising flux structures. We also note that there is a swift transition
as we vary our parameters between essentially passive structures and
ones that strongly disrupt the convection layer. We intend to
carry out a more extensive investigation of intermediate parameter
ranges in which the role of downward pumping is important in
counteracting the buoyancy effects \citep{TBC2}. Finally, we intend to
explore the interactions between shear-driven buoyancy instabilities
and convective flows at higher Richardson numbers (with a
hydrodynamically stable shear). This will be a challenging problem to
tackle numerically (requiring high numerical resolution), but results
from these preliminary calculations constitute a firm foundation for
future work.

\section*{Acknowledgements}

The authors thank Nic Brummell, Nigel Weiss and Geoff Vasil
for stimulating discussions.

The numerical calculations were carried out
on the UKMHD cluster based in St Andrews, which is partially funded by
STFC. This research is supported via a rolling grant from STFC that is
held at DAMTP, University of Cambridge. PJB and LJS also wish to
acknowledge support from the KITP, Santa Barbara and travel grants
from the RAS to facilitate attendance at a workshop where some of the
work  was done.

\label{lastpage}


\begin{thebibliography}{99}

\bibitem[\protect\citeauthoryear{Bushby \& Houghton}{2005}]{BH}
Bushby P. J., Houghton S. M., 2005, MNRAS, 362, 313.
\bibitem[\protect\citeauthoryear{Bushby}{2006}]{Bushby}
Bushby P. J., 2006, MNRAS, 371, 772.
\bibitem[\protect\citeauthoryear{Br\"uggen \& Hillebrandt}{2001}]{Bruggen}
Br\"uggen, M., Hillebrandt, W., 2001, MNRAS, 323, 56.
\bibitem[\protect\citeauthoryear{Brummell, Cline \& Cattaneo}{2002}]{BCC}
Brummell, N., Cline, K., Cattaneo, F., 2002, MNRAS, 329, L73.
\bibitem[\protect\citeauthoryear{Cattaneo \&
    Hughes}{1988}]{Catthughes}
Cattaneo F., Hughes D. W., 1988, JFM, 196, 323.
\bibitem[\protect\citeauthoryear{Cattaneo \& Hughes}{2006}]{CH2}
Cattaneo F., Hughes D. W., 2006, JFM, 553, 401.
\bibitem[\protect\citeauthoryear{Cattaneo, Brummell \& Cline}{2006}]{CBC}
Cattaneo F., Brummell N., Cline K. S., 2006, MNRAS, 365, 727.
\bibitem[\protect\citeauthoryear{Chan, Liao, Zhang \& Jones}{2004}]{Chan}
  Chan, K.\ H., Liao, X., Zhang, K., Jones, C. A., 2004, A\&A, 423,
  37.
\bibitem[\protect\citeauthoryear{Chandrasekhar}{1961}]{Chand}
Chandrasekhar, S.,  1961, Hydrodynamic and Hydromagnetic Stability,
Pub. Dover.
\bibitem[\protect\citeauthoryear{Charbonneau \&
    MacGregor}{1997}]{Char}
Charbonneau, P.\ \& MacGregor, K.\ B., 1997, ApJ, 486, 502.
\bibitem[\protect\citeauthoryear{Christensen-Dalsgaard \&
    Thompson}{2007}]{CT} Christensen-Dalsgaard, J.\ \&
    Thompson, M.\ J., 2007, Ch. 3 in The Solar Tachocline, CUP.
\bibitem[\protect\citeauthoryear{Dikpati \& Gilman}{2009}]{DG09}
Dikpati, M. \ \& Gilman, P.A., 2009, Space Science Reviews, DOI 10.1007/s11214-008-9484-3.
\bibitem[\protect\citeauthoryear{Dormy \& Soward}{2007}]{Dormy} Dormy
  E.\, Soward A.\ (eds), 2007, Mathematical Aspects of Natural
  Dynamos, CRC Press.
\bibitem[\protect\citeauthoryear{Frank, Jones, Ryu, Gaalaas}{1996}]{frank}
Frank, A., Jones, T. W., Ryu, D., Gaalaas, J. B., 1996, ApJ, 460, 777.
\bibitem[\protect\citeauthoryear{Fan}{2001}]{fan}
Fan Y., 2001, ApJ, 546, 509.
\bibitem[\protect\citeauthoryear{Hughes}{1985}]{Hughes1985} Hughes,
    D. W., 1985, GAFD, 32, 273.
\bibitem[\protect\citeauthoryear{Hughes \& Tobias}{2001}]{HT1} Hughes,
    D. W., Tobias, S. M., 2001, Proc. Roy. Soc. A., 457, 1365.
\bibitem[\protect\citeauthoryear{Hughes}{2007}]{Hughes2007} Hughes,
    D. W., 2007, The Solar Tachocline, pp 275--298 (CUP).
\bibitem[\protect\citeauthoryear{K\"apyl\"a, Korpi \&
    Brandenburg}{2008}]{Kap}
    K\"apyl\"a P. J., Korpi M. J., Brandenburg, A., 2008, A\&A, accepted.
\bibitem[\protect\citeauthoryear {Gilman}{1970}]{Gilman}
Gilman, P.A., 1970, ApJ, 162, 1019.
\bibitem[\protect\citeauthoryear {Gilman \& Cally}{2007}]{GC07}
Gilman, P.A. \& Cally, P.S., 2007, The Solar Tachocline, pp 243--274 (CUP).
\bibitem[\protect\citeauthoryear{Kersal\'e,  Hughes \& Tobias}{2007}]{Evy}
Kersal\'e E., Hughes D. W., Tobias S. M., 2007, ApJ, 663, L113
\bibitem[\protect\citeauthoryear{Jouve \& Brun}{2007}]{Jouve}
  Jouve, L., \& Brun, A.\ S., AN, 2007, 328, 10, 1104.
\bibitem[\protect\citeauthoryear{Lin, Silvers \& Proctor}{2008}]{Lin}
  Lin, M.-K. Silvers, L.\ J.\  \& Proctor, M.\ R.\ E., 2008,
  Phys. Lett. A, 373, 1, 69.
\bibitem[\protect\citeauthoryear{Matthews, Hughes \& Proctor}{1995}]{MHP}
Matthews P. C., Hughes D. W., Proctor M. R. E., 1995, ApJ, 448, 938.
\bibitem[\protect\citeauthoryear{Matthews, Proctor \& Weiss}{1995}]{MPW}
  Matthews P.\ C., Proctor M.\ R.\ E.\, Weiss N.\ O., JFM, 305, 281.
\bibitem[\protect\citeauthoryear{Moffatt}{1978}]{Moff}
Moffatt H. K., 1978, Magnetic field generation in electrically-conducting fluids, CUP: Cambridge.
\bibitem[\protect\citeauthoryear{Newcomb}{1961}]{Newcomb1961} Newcomb,
  W. A., 1961,  Phys. Fluids, 4 391.
\bibitem[\protect\citeauthoryear{Ossendrijver}{2003}]{Ossendrijver}
  Ossendrijver M.\, 2003, Astron. Astrophys. Rev., 11, 287.
\bibitem[\protect\citeauthoryear{Palotti, Heitsch, Zweibel, Huang}{2008}]{Palotti}
Palotti, M. L., Heitsch, F., Zweibel, E. G., Huang, Y.-M., 2008, ApJ, 678, 234.
\bibitem[\protect\citeauthoryear{Parker}{1955}]{Parker_buoy}
Parker E. N., 1955, ApJ, 121, 491.
\bibitem[\protect\citeauthoryear{Parker}{1993}]{Parker} Parker E.\
  N.\, 1993, ApJ, 408, 707.
\bibitem[\protect\citeauthoryear{Proctor}{2006}]{mp06} Proctor, M.R.E.\, 2006, EAS Pubs Series, 21, 241-273.
\bibitem[\protect\citeauthoryear{Ryu, Jones, Frank}{2000}]{ryu}
 Ryu, D., Jones, T. W., Frank, A. 2000, ApJ, 545, 475.
\bibitem[\protect\citeauthoryear{Silvers}{2008}]{silvers08}
  Silvers L. J, 2008, Phil. Roy. Trans. Soc. A.,  366, 4453.
\bibitem[\protect\citeauthoryear{Silvers {\it et al.}}{2009}]{siletal09} Silvers, L.J., Vasil, G.M., Brummell, N.C. \& Proctor, M.R.E., 2009, ApJL, submitted .
\bibitem[\protect\citeauthoryear{Spiegel \& Zahn}{1992}]{SZ}
Spiegel E. A., Zahn J.-P., 1992, Astron. Astrophys., 265, 106.
\bibitem[\protect\citeauthoryear{Tobias \textit{et al.}}{1998}]{TBCT1} Tobias S.\ M., Brummell N.\ H.,
  Clune T.\ L.\, Toomre, J., 1998, ApJL, 502, 177.
\bibitem[\protect\citeauthoryear{Tobias \textit{et al.}}{2001}]{TBC2} Tobias S.\ M., Brummell N.\ H.,
  Clune T.\ L., Toomre J., 2001, ApJL, 549, 2, 1183.
\bibitem[\protect\citeauthoryear{Tobias \& Hughes}{2004}]{HT2} Tobias
  S. M., Hughes, D. W., 2004, ApJ, 603, 785.
\bibitem[\protect\citeauthoryear {Tobias \& Weiss}{2007}]{TW07}
Tobias, S.M. \& Weiss, N.O., 2007, The Solar Tachocline, pp 319--350 (CUP).
\bibitem[\protect\citeauthoryear{Vasil \& Brummell}{2008}]{Vasila}
Vasil G. M., Brummell N. H., 2008, ApJ, 686, 709.
\bibitem[\protect\citeauthoryear{Vasil \& Brummell}{2009}]{Vasilb}
Vasil G. M., Brummell N. H., ApJ, 690, 783.

\bibitem[\protect\citeauthoryear{Wissink \textit{et al.}}{2000}]{Wissink} Wissink, J. G., Matthews, P. C., Hughes,
    D. W., Proctor, M. R. E., 2000, ApJ , 536, 2, 982-997.
\bibitem[\protect\citeauthoryear{Zhang, Liao, Schubert}{2004}]{Zhang}
  Zhang, K., Liao, X. Schubert, G., 2004, ApJ, 602, 468.





\end{thebibliography}
\end{document}